\def\simlt{\lower.5ex\hbox{$\; \buildrel < \over \sim \;$}}

\RequirePackage{lineno} 
\documentclass[preprint2]{emulateapj}
\usepackage[all,cmtip]{xy}
\usepackage{amsmath}
\usepackage[subnum]{cases}
\usepackage{mathtools}
\usepackage[english]{babel}
\usepackage{blindtext}
\usepackage{listings}
\usepackage{hyperref}
\allowdisplaybreaks
\newcommand*{\rom}[1]{\expandafter\@slowromancap\romannumeral #1@}
\newcommand{\myemail}{xiz@ucsc.edu}
\slugcomment{Accepted at ApJ}
\shorttitle{Vertical Tracer Mixing in Planetary Atmospheres Part I}
\shortauthors{Zhang \& Showman}
\begin{document}

\title{Global-mean Vertical Tracer Mixing in Planetary Atmospheres\\
I: Theory and Fast-rotating Planets}
\author{Xi Zhang$^1$ and Adam P. Showman$^2$}
\affil{$^1$Department of Earth and Planetary Sciences, University of California Santa Cruz, CA 95064}
\affil{$^2$Department of Planetary Sciences and Lunar and Planetary Laboratory, University of Arizona, AZ 85721}
\altaffiltext{1}{Correspondence to be directed to \myemail}

\begin{abstract}

Most chemistry and cloud formation models for planetary atmospheres adopt a one-dimensional (1D) diffusion approach to approximate the global-mean vertical tracer transport. The physical underpinning of the key parameter in this framework, eddy diffusivity $K_{zz}$, is usually obscure. Here we analytically and numerically investigate vertical tracer transport in a 3D stratified atmosphere and predict $K_{zz}$ as a function of the large-scale circulation strength, horizontal mixing due to eddies and waves and local tracer sources and sinks. We find that $K_{zz}$ increases with tracer chemical lifetime and circulation strength but decreases with horizontal eddy mixing efficiency. We demarcated three $K_{zz}$ regimes in planetary atmospheres. In the first regime where the tracer lifetime is short compared with the transport timescale and horizontal tracer distribution under chemical equilibrium ($\chi_0$) is uniformly distributed across the globe, global-mean vertical tracer mixing behaves diffusively. But the traditional assumption in current 1D models that all chemical species are transported via the same eddy diffusivity generally breaks down. We show that different chemical species in a single atmosphere should in principle have different eddy diffusion profiles. In the second regime where tracer is short-lived but $\chi_0$ is non-uniformly distributed, a significant non-diffusive component might lead to a negative $K_{zz}$ under the diffusive assumption. In the third regime where the tracer is long-lived, global-mean vertical tracer transport is also largely influenced by non-diffusive effects. Numerical simulations of 2D tracer transport on fast-rotating zonally symmetric planets validate our analytical $K_{zz}$ theory over a wide parameter space.

\end{abstract}


\section{Introduction}

The spatial patterns of atmospheric tracers such as chemical species, haze and cloud particles are significantly influenced by atmospheric transport. For most planets in the solar system, the horizontal variations of tracers are usually much smaller than their vertical variations. For extra-solar planets, the horizontal distributions of tracers are not well resolved by observations. The global-mean vertical tracer distribution is the primary focus in most chemistry and cloud models. Vertical transport by large-scale atmospheric circulation and wave breaking leads to a tracer profile that deviates from its chemical or microphysical equilibrium. One example is the ``quenching" phenomenon in the atmospheres of giant planets and close-in exoplanets where the gas tracer concentrations (e.g., CO on Jupiter, \citealt{prinn1977carbon}) in the upper atmosphere are not in their local chemical equilibrium, but instead are dominated by strong vertical tracer mixing from the deeper atmosphere. As a result, the observed tracer abundance in the upper atmosphere is ``quenched" at an abundance corresponding to the chemical-equilibrium abundance at the location where the tracer chemical timescale is equal to the vertical mixing timescale. This disequilibrium ``quenching effect" has a significant influence on the interpretation of observed spectra of planetary atmospheres (e.g., \citealt{prinn1976chemistry}, \citealt{prinn1977carbon}, \citealt{smith1998estimation}, \citealt{cooper-showman-2006}, \citealt{visscher-moses-2011}).

Investigation of global-mean vertical tracer distribution could be simplified to a one-dimensional (1D) problem. However, because atmospheric tracer transport is intrinsically a three-dimensional (3D) process, parameterization of the 3D transport in a 1D vertical transport model is tricky. Most chemistry models for solar system planets (e.g., \citealt{allen1981vertical}, \citealt{krasnopolsky1981chemical}, \citealt{yung1984photochemistry}, \citealt{nair1994photochemical}, \citealt{moses2005photochemistry}, \citealt{lavvas2008coupling}, \citealt{zhang2012sulfur}, \citealt{wong2017photochemistry}, \citealt{moses2017dust}) and exoplanets (e.g., \citealt{moses-etal-2011}, \citealt{line2011thermochemical}, \citealt{hu2012photochemistry}, \citealt{tsai2017vulcan}) adopt a 1D chemical-diffusion approach. Most haze and cloud formation models on planets and brown dwarfs assume the vertical particle transport behaves in a diffusive fashion as well (e.g., \citealt{turco1979one}, \citealt{ackerman-marley-2001}, \citealt{helling2008dust}, \citealt{gao2014bimodal}, \citealt{lavvas2017aerosol}, \citealt{gao2017pluto}, \citealt{powell2018formation}). This type of 1D framework originated in the Earth atmospheric chemistry literature but gradually faded out since the 1990's because it is not sufficient to simultaneously explain the distributions of all the species from observations (\citealt{holton1986dynamically}) and when the 3D distributions of multiple tracers are well determined by satellite observations from space. At the moment, this 1D framework is still very useful in planetary atmospheric studies. Although only a crude approximation, 1D models have succeeded in explaining the observed global-averaged vertical profiles of many chemical species and particles on planets (e.g., Titan by \citealt{yung1984photochemistry} and  \citealt{zhang2010atomic}, Jupiter by \citealt{moses2005photochemistry}, Venus by \citealt{zhang2010venus}, \citealt{zhang2012sulfur} and \citealt{gao2014bimodal}, Pluto by \citealt{wong2017photochemistry} and  \citealt{gao2017pluto}).

\begin{figure}[t]
  \centering \includegraphics[width=0.48\textwidth]{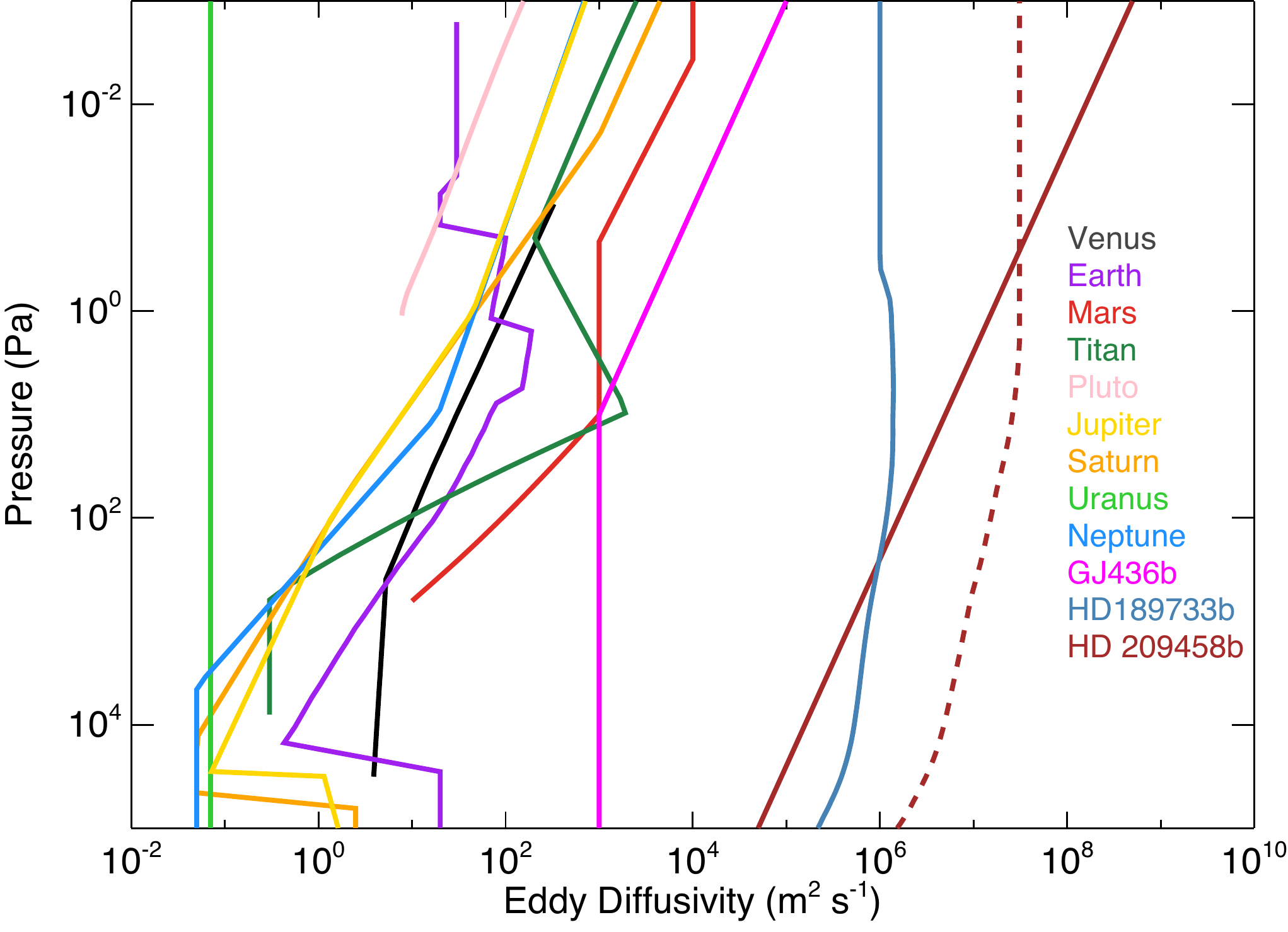} 
  \caption{Vertical profiles of eddy diffusivity in typical 1D chemical models of planets in and out of the solar system. Sources:  \citet{zhang2012sulfur} for Venus, \citet{allen1981vertical} for Earth, \citet{nair1994photochemical} for Mars, \citet{li2015vertical} for Titan, \citet{wong2017photochemistry} for Pluto, \citet{moses2005photochemistry} for Jupiter, Saturn, Uranus and Neptune, \citet{moses2013compositional} for GJ436b and \citet{moses-etal-2011} for HD189733b and HD209458b. For HD209458b, we show eddy diffusivity profiles assumed in a gas chemistry model (dashed, \citealt{moses-etal-2011}) and that derived from a 3D particulate tracer transport model (solid, \citealt{parmentier20133d}).} 
\end{figure}

In the 1D chemical-diffusion framework, the strength of the vertical diffusion is characterized by a parameter called eddy diffusivity or eddy mixing coefficient. The 1D effective eddy diffusivity $K_{zz}$ is often determined empirically by fitting the observed vertical tracer profiles. Fig. 1 shows that $K_{zz}$ on different planets could exhibit vertical profiles and magnitude varying across more than eight orders of magnitude. This variation implies different vertical transport efficiencies from planet to planet. Since all 3D dynamical effects have been lumped into a single eddy diffusivity, the specific dynamical mixing mechanisms that lead to a particular vertical profile of $K_{zz}$ are often obscure. If the atmosphere is convective, then using the traditional Prandtl mixing length theory (e.g., \citealt{prandtl1925bericht}, \citealt{smith1998estimation}, \citealt{bordwell2018convective}), one can formulate $K_{zz}$ as a product of a convective velocity and a typical vertical length scale in a turbulent medium. But this formalism fails when the atmosphere is stably stratified. In the low-density middle and upper atmosphere such as Earth's mesosphere, the vertically propagating gravity waves could break and also lead to a strong vertical mixing of the chemical tracers. \citet{lindzen1981turbulence} parameterized the eddy diffusivity from the turbulence and stress generated in breaking gravity and tidal waves (also see discussion in \citealt{strobel1981parameterization}, \citealt{strobel1987vertical}). In a stratified atmosphere such as Earth's stratosphere, tracer transport is subjected to both large-scale overturning circulation and vertical wave mixing (\citealt{hunten1975vertical}, \citealt{holton1986dynamically}), but their relative importance depends on altitude and many other factors and may differ from planet to planet. 

One of the conventional assumptions in the existing framework used in current planetary models is that all tracers, in spite of their different chemical lifetimes or particle microphysical/settling timescales, are simulated using the same eddy diffusivity profile $K_{zz}$. The tracer distribution in the real atmosphere is controlled by 3D dynamical and chemical/microphysical processes. Therefore a coupling feedback between the chemistry and vertical transport is expected. Actually it has been noticed in the Earth community (e.g., \citealt{holton1986dynamically}), in the presence of meridional (latitudinal) transport in the stratospheres, the derived effective eddy diffusivity as a global-mean transport coefficient could have a strong dependence on the tracer lifetime, and thus its chemical/microphysical sources and sinks. Although there are several 3D chemical-transport simulations in planetary atmospheres with simplified chemical and cloud schemes (e.g., \citealt{lefevre2004three}, \citealt{lefevre2008heterogeneous}, \citealt{marcq2013simulations}, \citealt{stolzenbach2015three}, \citealt{cooper-showman-2006}, \citealt{parmentier20133d}, \citealt{charnay20153d2}, \citealt{lee2016dynamic}, \citealt{drummond2018effect}, \citealt{lines2018simulating}), a thorough understanding of the physical basis of global-mean vertical tracer transport and $K_{zz}$ using both analytical theory and 2D and 3D numerical simulations is still lacking.

Here we aim to reexamine this conventional 1D diffusion framework. We wish to achieve a more physically based parameterization of $K_{zz}$ from first principles. Our study will be presented in two consecutive papers. In Paper I (the current paper), we will construct a first-principles theory of $K_{zz}$ in a 3D atmosphere and numerically investigate the behaviors of $K_{zz}$ on fast-rotating planets using a 2D chemical-transport model. In Paper II (\citealt{zhang2018kzz}), we will specifically focus on 3D chemical tracer transport on tidally locked exoplanets and the associated $K_{zz}$ using a 3D general circulation model (GCM). We primarily focus on stratified atmospheres such as the stratosphere on solar-system planets or the photospheres on highly irradiated exoplanets where most of the chemical tracers and haze/clouds are observed, and which are expected to be stably stratified (e.g., \citealt{fortney2008unified}, \citealt{madhusudhan2009temperature}, \citealt{line2012information}). 

In the following sections of this paper, we will first elaborate the underlying physics of global-mean tracer transport and construct a first-principles estimate of $K_{zz}$. Then we will use a 2D chemical-transport model to study a 2D meridional circulation system on fast-rotating planets and the behaviors of the associated $K_{zz}$ under several typical scenarios and $K_{zz}$ regimes. We conclude this study with several key statements and a brief discussion on the effect of vertically propagating gravity waves on the vertical transport of tracers.

\section{Theoretical Background}
\subsection{Nature of the problem}

In a stratified atmosphere, tracers tend to be mixed upward by the large-scale overturning circulation if there is a correlation on an isobar between tracer abundance and vertical velocity: if the tracer abundance is high where the vertical velocity is upward, or if the tracer abundance is low where the vertical velocity is downward, tracer is mixed upward (Fig. 2). This implies that the net vertical mixing of tracer over the globe depends crucially on horizontal variations of the tracer on isobars and on their correlation with the vertical velocity field (\citealt{holton1986dynamically}). If the tracer distribution is initially horizontally uniform across the globe with a vertical gradient of the mean tracer abundance, vertical wind transport will naturally produce tracer perturbations on an isobar that are correlated with the vertical velocity field (Fig. 2). For example, if the initial tracer mixing ratio is higher in the lower atmosphere and lower in the upper atmosphere, the upwelling branch of the overturning circulation will transport the higher-mixing-ratio tracers upward and the downwelling branch will transport the lower-mixing-ratio tracers downward from the upper atmosphere. As a result, the tracers on an isobar will be more abundant in the upwelling branch than in the downwelling branch, exhibiting a positive correlation with the vertical velocity field (Fig. 2). After this horizontal tracer distribution is established, the upwelling branch will transport higher-mixing-ratio tracers across an isobar and the downwelling branch will transport lower-mixing-ratio tracers across the same pressure level---the upward and downward tracer fluxes do not cancel out. When averaged over the globe, there will be a net upward tracer flux, resulting in an effective upward tracer transport in the global-mean sense. 

Several other processes act to enhance or damp those horizontal tracer perturbations on the isobar. Horizontal mixing/diffusion due to eddies and waves (e.g., breaking of Rossby waves) normally smooth out the tracer variation across the globe (e.g., \citealt{holton1986dynamically},  \citealt{yung2009evidence},  \citealt{friedson2012general}). Horizontal advection due to the mean flow may increase or decrease the horizontal tracer variations, depending on the correlation between the horizontal velocity convergence/divergence and the tracer distribution on the isobar. The distribution of tracer sources and sinks due to non-dynamical processes such as chemistry\footnote{Hereafter we just use the generic term ``chemistry" to represent any non-dynamical processes that affect the tracer distribution, such as chemical reactions in the gas and particle phase, haze and cloud formation, or other phase transition processes.} also plays an important role. In a simplified picture, those non-dynamical processes could be assumed to relax the tracer distribution back to the chemical equilibrium distribution that the tracers would have in the absence of dynamics, which could be either uniformly distributed across the globe or with a significant variations depending on the local sources and sinks (\citealt{marcq2013simulations}). It is expected that the 1D effective eddy diffusivity $K_{zz}$ depends on the magnitude of the vertical velocity, chemical timescale of the species, horizontal transport timescale and the horizontal variation of the tracer distribution under chemical equilibrium.

\begin{figure}[t]
  \centering \includegraphics[width=0.48\textwidth]{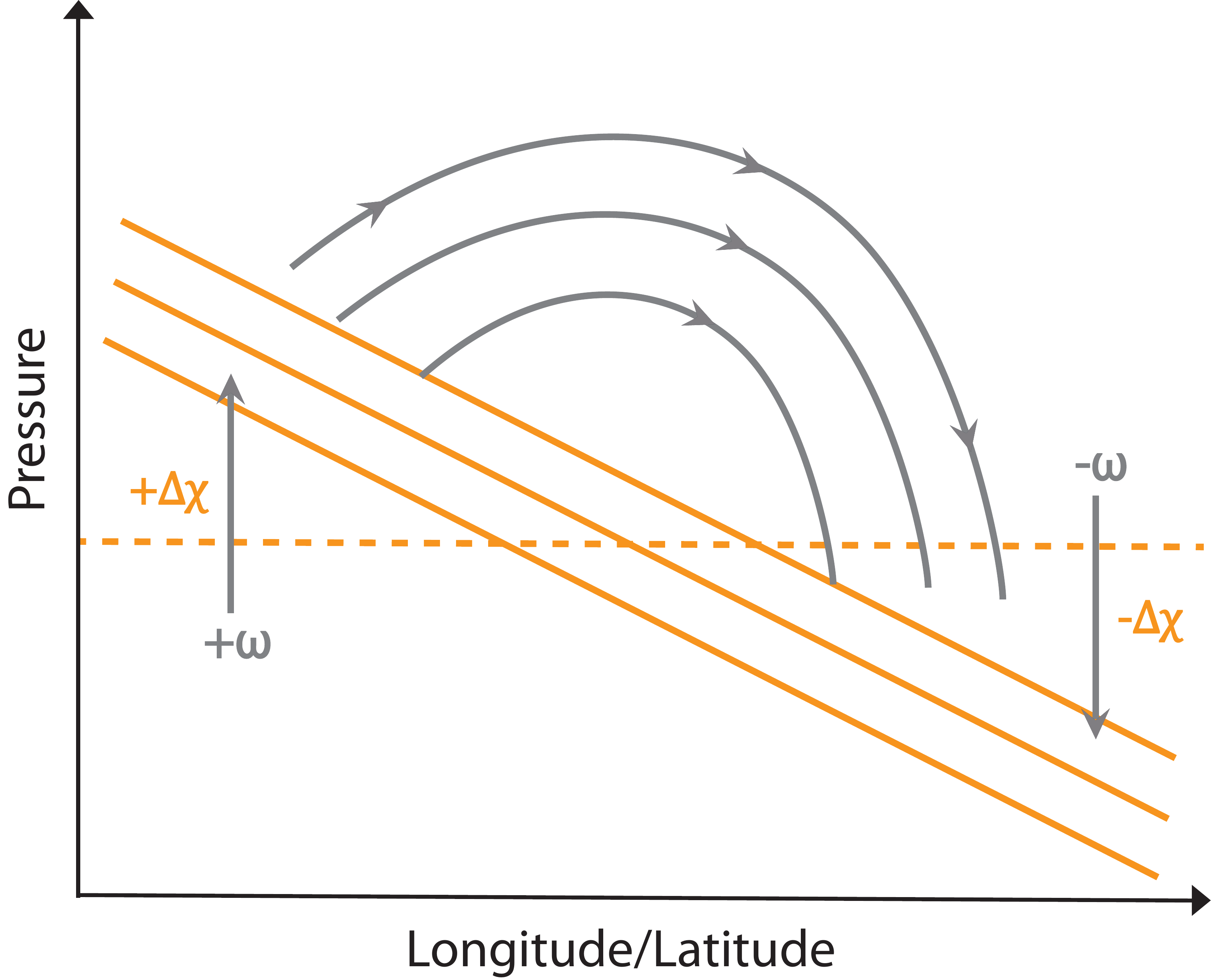} 
  \caption{Illustration of tracer transport with a large-scale circulation. The orange solid lines indicate constant tracer mixing ratio surfaces. Tracer mixing ratio is higher in the lower atmosphere. Gray lines show horizontal and vertical wind transport with vertical velocity $w$. $\Delta\chi$ is the deviation of local tracer mixing ratio from the horizontally averaged mixing ratio on an isobar (dashed).} 
\end{figure}

\citet{holton1986dynamically} first quantified these effects in the Earth's atmosphere with a scenario that envisions the vertical mixing is accomplished by a meridional circulation in a 2D (latitude-pressure) framework. The derived eddy diffusivity exhibits a strong dependence on the circulation strength, tracer chemical lifetime and horizontal mixing. Holton showed that the species-dependent eddy diffusivity might help simultaneously explain the vertical profiles of several species in the stratosphere of Earth, whereas the species-independent eddy diffusivity could not. Here we generalize the 2D theory from \citet{holton1986dynamically} to a 3D atmosphere so that it can also be applied to other planets that are not as zonally symmetric as the Earth. 

\subsection{Governing Equation of Vertical Tracer Transport}

Here we include the tracer advection by 3D atmospheric dynamics and tracer chemistry via a simplified chemical scheme to study the global-mean tracer transport. Based on these investigations we will achieve an analytical parameterization of the 1D effective eddy diffusivity $K_{zz}$. We will also demarcate different atmospheric regimes in terms of the tracer chemical lifetime and horizontal tracer distribution under chemical equilibrium. 

First we start from a general 3D tracer transport equation:
\begin{eqnarray}
\frac{D\chi}{Dt}=S
\end{eqnarray}
where $S$ is the net sources/sinks of the chemical tracer with mixing ratio $\chi$. In principle, isentropic coordinates are more appropriate for discussion of the tracer transport (\citealt{AHL1987}). But for simplicity, here we just adopt the log-pressure coordinate $\{x,y,z\}$. The coordinates are defined as $x=a\lambda\sin{\phi}$ and $y=a\phi$, where $a$ is the planetary radius, $\lambda$ is longitude and $\phi$ is latitude. Here, $z \equiv -H\log (p/p_0)$ is the log-pressure where $H$ is a constant reference scale height, $p$ is pressure and $p_0$ is the reference pressure in the pressure coordinate. $D/Dt=\partial/\partial{t} + \vec{\bf{u}}\cdot\nabla_h+w\partial/\partial{z}$ is the material derivative. $\nabla_h=(\partial/\partial{x}, \partial/\partial{y})$ is the horizontal gradient at constant pressure. $\vec{\bf{u}}=(u, v)$ is the horizontal velocity at constant pressure where $u$ is the zonal (east-west) velocity and $v$ is the meridional (north-south) velocity. $w=Dz/Dt$ is the vertical velocity.  In the log-pressure coordinates, Eq. (1) becomes:
\begin{eqnarray}
\frac{\partial \chi}{\partial t}+\vec{\bf{u}}\cdot\nabla_h\chi+w\frac{\partial \chi}{\partial z}=S.
\end{eqnarray}

The continuity equation is:
\begin{eqnarray}
\nabla_h \cdot \vec{\bf{u}}+e^{z/H}\frac{\partial}{\partial z}(e^{-z/H}w)=0.
\end{eqnarray}

Combining the above two equations, we get the flux form of the tracer-transport equation:
\begin{eqnarray}
\frac{\partial \chi}{\partial t}+\nabla_h \cdot (\chi\vec{\bf{u}})+e^{z/H}\frac{\partial}{\partial z}(e^{-z/H}w\chi)=S.
\end{eqnarray}

Here we define an eddy-mean decomposition $A=\overline{A}+A^\prime$ where A represents any quantity. $\overline{A}$ is the globally average quantity at constant pressure and $A^\prime$ is the deviation from the mean, or the ``eddy" term. Taking the global average of the continuity Eq. (3) to eliminate the the horizontal divergence term, we obtain $\overline{w}=0$ for each atmospheric level if we assume the global-mean vertical velocity $\overline{w}$ vanishes at top and bottom boundaries. Globally averaging Eq. (4) and using $\overline{w}=0$, we obtain:
\begin{eqnarray}
\frac{\partial\overline{\chi}}{\partial t}+e^{z/H}\frac{\partial}{\partial z}(e^{-z/H}\overline{w\chi^\prime})=\overline{S}.
\end{eqnarray}

This is the global-mean vertical tracer transport equation. It states that the evolution of global-mean tracer mixing ratio $\overline{\chi}$ is related to its global-mean vertical eddy fluxes. Based on Eq. (5), $\overline{\chi}$ cannot be solved directly unless we establish a relationship between the mean value and the eddy flux $\overline{w\chi'}$. A conventional assumption is the ``flux-gradient relationship'' that links the eddy tracer flux to the vertical gradient of the mean value (e.g. Plumb and Mahlman 1987) by introducing a 1D ``effective eddy diffusion'' $K_{zz}$ such that:
\begin{eqnarray}
\overline{w\chi^\prime}\approx -K_{zz}\frac{\partial\overline{\chi}}{\partial z}.
\end{eqnarray}

If Eq. (6) is valid, the global-mean tracer transport equation (5) can be formulated as a vertical diffusion equation: 
\begin{eqnarray}
\frac{\partial\overline{\chi}}{\partial t}-e^{z/H}\frac{\partial}{\partial z}(e^{-z/H}K_{zz}\frac{\partial\overline{\chi}}{\partial z})=\overline{S}.
\end{eqnarray} 

This is the widely-used 1D chemical-diffusion equation. We need to estimate the eddy flux in Eq. (6) to solve for $K_{zz}$. Subtracting both Eq. (5) and Eq. (3) that is multiplied by $\overline{\chi}$ from Eq. (4), we obtain:
\begin{equation}
\frac{\partial\chi^\prime}{\partial t}+\nabla_h \cdot (\chi^\prime\vec{\bf{u}})+w\frac{\partial\overline{\chi}}{\partial z}+e^{z/H}\frac{\partial}{\partial z}[e^{-z/H}(w\chi^\prime-\overline{w\chi^\prime})]=S^\prime.
\end{equation}

Now we need to estimate the horizontal tracer variation $\chi^\prime$ along an isobar. In terms of the tracer lifetime $\tau_c$ and horizontal tracer distribution under chemical equilibrium $\chi_{0}$, the behavior of $K_{zz}$ can be categorized into three typical regimes  (Fig. 3): (I) a short-lived tracer with uniform distribution of chemical equilibrium abundance, (II) a short-lived tracer tracer with non-uniform distribution of chemical equilibrium abundance, and (III) a long-lived tracer whose lifetime is long compared with the transport timescale.

\begin{figure}[t]
\includegraphics[width=0.48\textwidth]{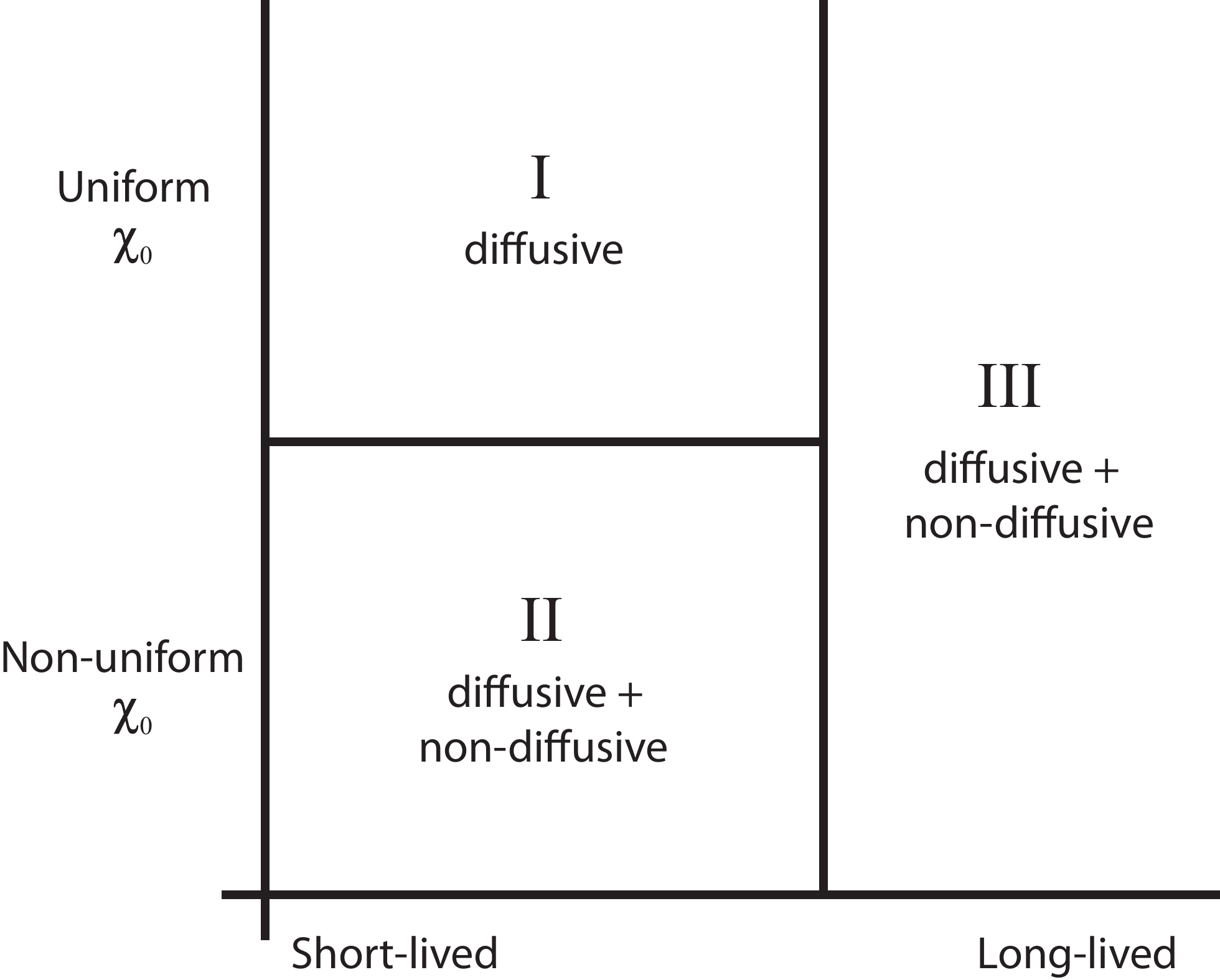} 
 \centering\caption{Atmospheric regimes of global-mean tracer transport as a function of tracer chemical lifetime and horizontal tracer distribution under chemical equilibrium $\chi_0$.} 
\end{figure}

\subsection{Regime I: Short-lived Tracers with Uniform Chemical Equilibrium: Diffusive case}

For short-lived tracers, we can estimate $\chi^\prime$ from the vertical gradient of $\overline{\chi}$ under following four assumptions.

(i) We neglect the temporal variation (time evolution) of the $\chi^\prime$ in statistical steady state because we focus on the time-averaged behavior in this study. The first term on the left hand side of Eq. (8) can be neglected.

(ii) We neglect the complicated eddy term (the last term in the left hand side of Eq. (8)) that could be much smaller than the third term. In other words, we assume that the deviation of the eddy flux from its global mean is smaller than the local vertical transport of the global-mean tracer. As shown later in numerical simulations, this is generally valid in the case where the deviation of the tracer mixing ratio from the mean is small and the material surface is not significantly distorted ($\chi^\prime$ is small), or where the vertical gradient of $\chi^\prime$ is small compared with the mean tracer gradient. For long-lived tracers (Regime III), this assumption is not valid.

(iii) We approximate the horizontal tracer eddy flux term $\nabla_h \cdot (\chi^\prime\vec{\bf{u}})\approx\chi^\prime/\tau_{d}$ where $\tau_{d}$ is the characteristic dynamical timescale in the horizontal mixing processes. In general, under different situations, this divergence term could act in an advective way, or in a diffusive way, motivating two possible ways of formulating the dynamical timescale. In the advection, the dynamical transport timescale $\tau_{d}\approx L_h/U_{adv}$ where $L_h$ is horizontal characteristic length scale and $U_{adv}$ is the horizontal wind speed. In the diffusive case, the dynamical timescale $\tau_{d}\approx L_h^2/D$ where $D$ is the effective horizontal eddy diffusivity. The horizontal length scale $L_h$ is determined by dominant flow patterns in the atmosphere (see detailed discussions on p.9 in  \citealt{perez-becker-showman-2013}). For tracers transported by a global-scale circulation pattern (e.g., equator-to-pole meridional circulation), $L_h$ is usually taken as the planetary radius $a$. Note that in this linear relaxation approximation, we have assumed that horizontal dynamics always reduces the horizontal variation of the tracer. This generally holds true if the horizontal tracer field is not complicated. Some exceptions will be discussed in the numerical simulation sections later. 

(iv) For simplicity, we consider a linear chemical scheme which relaxes the tracer distribution towards local chemical equilibrium $\chi_0$ in a timescale $\tau_c$:
\begin{eqnarray}
S=\frac{\chi_0-\chi}{\tau_{c}}.
\end{eqnarray}

In general, the chemical equilibrium tracer distribution $\chi_0$ depends on many local factors, such as temperature, abundances of other species, photon fluxes and precipitating ion fluxes. It is expected that photochemical species will exhibit $\chi_0$ that varies between the equator and poles. These effects might have more pronounced influences on the horizontal distributions of $\chi_0$ on tidally locked exoplanets than on solar system planets. If the species advected upward from the deep atmosphere with a uniform thermochemical source, $\chi_0$ is assumed constant along an isobar. Without losing generality, here we consider a non-uniform horizontal distribution of the chemical equilibrium: $\chi_0=\overline{\chi_0}+\chi_0^{\prime}$, where $\overline{\chi_0}$ is the global-mean of $\chi_0$ which is only a function of pressure, and $\chi_0^{\prime}$ (which can be a function of longitude and latitude as well as pressure) is the departure of equilibrium tracer abundance from its global mean. Inserting $\chi_0$ into Eq. (9), we obtain the globally averaged chemical source/sink term $\overline{S} = (\overline{\chi_0}-\overline\chi)/\tau_c$ and the departure $S^\prime=(\chi_0^{\prime}-\chi^\prime)/\tau_c$.

With the assumptions (i-iv), Eq. (8) can be written:
\begin{equation}
w\frac{\partial\overline{\chi}}{\partial z}+\frac{\chi^\prime}{\tau_{d}}=\frac{\chi_0^{\prime}-\chi^\prime}{\tau_{c}}.
\end{equation}

We solve for $\chi^\prime$:
\begin{equation}
\chi^\prime=\frac{-w\frac{\partial\overline{\chi}}{\partial z}+\tau_{c}^{-1}\chi_0^{\prime}}{\tau_{d}^{-1}+\tau_{c}^{-1}}.
\end{equation}

This expression for the 3D distribution of $\chi'$ is qualitatively similar to the previous 2D model result in \citet{holton1986dynamically} with a photochemical source (c.f., his Eq. 19). However, \citet{holton1986dynamically} mainly focused on a special case ($\chi_0^{\prime}=0$) without further elaborating the physical meaning of the general expression Eq. (11). Here we explicitly point out that the global-mean vertical tracer transport is composed of two physical processes: a diffusion process and a non-diffusive process. Based on Eq. (11), the global-mean vertical tracer flux $\overline{w\chi^\prime}$ is:
\begin{equation}
\overline{w\chi^\prime}=\frac{-\overline{w^2}}{\tau_{d}^{-1}+\tau_{c}^{-1}}\frac{\partial\overline{\chi}}{\partial z}+\frac{\overline{w\chi_0^{\prime}}}{1+\tau_d^{-1}\tau_{c}}.
\end{equation}

The first term in Eq. (12) implies diffusive behavior, because this term's contribution to the tracer vertical flux is proportional to the vertical gradient of the mean tracer abundance. But the second term does not depend on the mean tracer gradient, suggesting a non-diffusive behavior. Instead, the second term originates from the correlation between the equilibrium tracer distribution and the vertical wind field. It can be neglected if the tracer under chemical equilibrium is more or less uniformly distributed across the globe. However, if the equilibrium tracer distribution is significantly non-uniform (for instance, if, in chemical equilibrium, the equator and poles have strongly differing chemical abundances), the conventional ``eddy diffusion" framework breaks down because the non-diffusive process might dominate the vertical tracer transport in the global-mean sense. To elaborate the physics here, we now further discuss these two cases: a uniformly distributed tracer under chemical equilibrium and a non-uniform case. 

If the equilibrium tracer distribution is uniformly distributed (i.e., $\chi_0^{\prime}=0$), the non-diffusive term in Eq. (12) vanishes. In this special case of short chemical lifetime (among other assumptions), mean tracer transport can be regarded as a diffusive process and the eddy diffusivity $K_{zz}$ can be parameterized based on the relationship between eddy tracer flux to the global-mean tracer gradient based on Eq. (12) and Eq. (6):
\begin{equation}
K_{zz}\approx\frac{\overline{w^2}}{\tau_{d}^{-1}+\tau_{c}^{-1}}.
\end{equation}

This result is also consistent with that from a zonal-mean 2D model in \citet{holton1986dynamically} with a uniform chemical equilibrium abundance. The parameterized $K_{zz}$ depends on circulation strength, chemical lifetime of the tracer and horizontal transport/mixing timescale in the atmosphere. It is expected that different chemical tracers are subjected to different eddy diffusivity strength, which has not been considered to date in 1D chemical models of planetary atmospheres.

If $\chi_0^{\prime}=0$, the horizontal distribution of $\chi^\prime$, deviation of the tracer mixing ratio from the mean on an isobar, is correlated with the horizontal distribution of vertical velocity $w$ at the same pressure level (Eq. 11):
\begin{equation}
\chi^\prime=\frac{-w\frac{\partial\overline{\chi}}{\partial z}}{\tau_{d}^{-1}+\tau_{c}^{-1}}.
\end{equation}

If the tracer has a large, uniform chemical equilibrium abundance in the deep atmosphere but is photochemically destroyed in the upper atmosphere---for example, sulfur dioxide on Venus (\citealt{zhang2012sulfur}), methane on Jupiter (\citealt{moses2005photochemistry}) or water on hot Jupiters (\citealt{moses-etal-2011})---the mean tracer gradient $\partial\overline{\chi}/\partial z$ is negative, and therefore $\chi^\prime$ is positively correlated with $w$ (Eq. 14). This implies that $\overline{w \chi'}>0$, i.e., the tracer is transported upward. On the other hand, if the tracer source is in the upper atmosphere, for example, chemically produced species which are uniformly distributed under chemical equilibrium, or oxygen species with uniform fluxes from comets into the upper atmospheres of giant planets (\citealt{moses2017dust}), the mean tracer gradient $\partial\overline{\chi}/\partial z$ is generally positive and $\chi^\prime$ is anti-correlated with $w$. This implies that $\overline{w \chi'}<0$, i.e., the tracer is transported downward. In both cases, the $\chi^\prime-w$ correlation patterns result in a net tracer transport away from their source regions. 

From Eq. (14), if the circulation is stronger, i.e., $w$ is larger, the horizontal variation of the tracer mixing ratio will be larger. If the chemical loss or the horizontal mixing due to advection or diffusion is stronger, i.e., $\tau_d$ and $\tau_c$ are smaller, the tracer variation is smaller. In other words, a large-scale circulation enhances the tracer variation on isobars, whereas chemical relaxation and horizontal tracer mixing homogenizes the tracer distribution on isobars. The resulting eddy diffusivity is larger if the circulation is stronger, the tracer chemical lifetime is longer, or the horizontal mixing timescale is longer.

We emphasize that $K_{zz}$ should depend on the chemical lifetime $\tau_c$---an important relationship that all current 1D models have ignored. To further elaborate this, we consider two different tracers, one with short $\tau_c$ and another with long $\tau_c$, that both exist in an atmosphere with a specified 3D atmospheric circulation. For concreteness, imagine that the background tracer abundance of both tracers decreases upward, with the same background gradient for both tracers. Because of the advection, for both tracers, the tracer abundance on an isobar will be greater in upwelling regions and smaller in downwelling regions, implying an upward flux of the tracer in both cases. However, when $\tau_c$ is short, the chemistry very strongly relaxes the abundance toward equilibrium, whereas with long $\tau_c$, this relaxation is weak. This implies that, in statistical equilibrium, the deviation of the actual tracer abundance from chemical equilibrium, $\chi'$, is greater when $\tau_c$ is long than when it is short.  Since the circulation is the same in the two cases, the upward tracer flux $\overline{w\chi^{\prime}}$ is greater when $\tau_c$ is long and smaller when $\tau_c$ is short.  Given the definition of eddy diffusivity $K_{zz}=-\overline{w\chi^{\prime}}/{\partial\overline{\chi}\over \partial z}$ from Eq. (6), we thus have the situation where the tracer with short $\tau_c$ has a smaller $K_{zz}$ than the tracer with large $\tau_c$---even though the atmospheric circulation, by definition, is precisely the same for the two tracers.  Indeed, for this simple situation, the value of $K_{zz}$ would go to zero in the limit $\tau_c\rightarrow 0$, because in that case, the tracer is constant on isobars, so there is no net correlation between vertical velocity and tracer abundance, implying that the upward flux of tracer is zero.

\subsection{Regime II: Short-lived Tracer with Non-uniform Chemical Equilibrium: Non-diffusive Component}

The horizontal distribution of tracer equilibrium abundance could be significantly non-uniform, i.e., $\chi_0^{\prime}\neq0$. This could occur if there is a local plume source of the species, for example, volcanic eruption on Earth (\citealt{self1993atmospheric}), convective injection of sulfur species to the middle atmosphere of Venus (\citealt{marcq2013variations}), and impact debris from incoming comets on Jupiter (\citealt{friedson1999transport}). A more common case is photochemically produced species like ozone on terrestrial planets or ethane on giant planets, where the incoming ultraviolet solar flux changes with the solar angle, leading to different photochemical equilibrium abundances at different latitudes (e.g., \citealt{moses2005latitudinal}). An extreme example is the tidally locked exoplanets, on which different chemical equilibrium states are expected between the permanent dayside and nightside for a tracer whose chemistry critically depends on factors such as temperature, incoming photon and ion fluxes. For example, formation of condensed haze/cloud particles favor the colder nightside than the warmer dayside (\citealt{powell2018formation}). For another example, species produced by photochemistry or ion chemistry are expected to have a larger equilibrium abundance on the dayside than on the nightside. 

In these situations, the dependence of $K_{zz}$ on the circulation and tracer chemistry is more complicated. The eddy tracer flux due to the non-diffusive process, i.e., the second term in the right hand side of Eq. (12), cannot be neglected. The horizontal distribution of $\chi^\prime$ might not be strongly correlated with the horizontal distribution of vertical velocity $w$ at the same pressure level (Eq. 11). The correlation term $\overline{w \chi_0^{\prime}}$ could have a significant effect on $K_{zz}$ (Eq. 12). Physically speaking, in the case with uniform chemical equilibrium abundance, atmosphere circulation will shape the initially homogeneous tracer distribution toward a correlation pattern that $\overline{w\chi^\prime}$ has the opposite sign as the background gradient $\partial\overline{\chi}/\partial z$. This implies that the tracer is mixed down the vertical gradient of the horizontal-mean tracer abundance. In the case of non-uniform chemical equilibrium case, $\chi_0$ might correlate or anti-correlate with the vertical velocity pattern, causing an additional contribution to the net vertical tracer transport in an essentially non-diffusive way. Whether this contribution will enhance or reduce the net vertical mixing efficiency depends on the correlation as well as the vertical gradient of the mean tracer (Eq. 12). 

If we still adopt the traditional ``eddy diffusion" framework, inserting the global-mean vertical tracer flux $\overline{w\chi^\prime}$ (Eq. 12) into the definition of $K_{zz}$ (the flux-gradient relationship Eq. 6), we can approximate the non-diffusive behavior and estimate $K_{zz}$ for tracers with non-uniform chemical equilibrium:
\begin{equation}
K_{zz}\approx\frac{\overline{w^2}}{\tau_{d}^{-1}+\tau_{c}^{-1}}-\frac{\overline{w\chi_0^{\prime}}}{1+\tau_{d}^{-1}\tau_{c}}(\frac{\partial\overline{\chi}}{\partial z})^{-1}.
\end{equation}

The second term on the right hand side represents inherently non-diffusive behavior, because it has a dependence on the vertical gradient of the mean tracer $\partial\overline{\chi}/\partial z$.  This implies that the eddy tracer flux (Eq. 6) has a term that does not scale linearly with the background vertical tracer gradient, which violates the fundamental assumption of a diffusive system that the flux scales linearly with the tracer gradient.  Also, unlike the first term in the right hand side of Eq. (15), the second term can have either sign, depending on the sign of the correlation between $w$ and the anomalies on isobars of the chemical-equilibrium abundance, $\chi'_0$. Note that Eq. (15) is no longer a closed expression for $K_{zz}$ because it depends on the vertical gradient of the global-mean tracer mixing ratio, a quantity that we require $K_{zz}$ to solve for. One could imagine that an iterative process between the chemical-diffusion simulation and updating $K_{zz}$ might lead to a final steady state of the system.

We emphasize that the $K_{zz}$ expression with the non-diffusive correction needs to be used cautiously in 1D chemical-diffusion simulations and can only be used when the non-diffusive contribution is not dominant. If the second term dominates and is negative, the predicted $K_{zz}$ can be negative in some situation. For example, if the tracer lifetime is very short, $\tau_c\rightarrow 0$, Eq. (15) becomes: 
\begin{equation}
K_{zz}\approx-\overline{w\chi_0^{\prime}}(\frac{\partial\overline{\chi}}{\partial z})^{-1}
\end{equation}

In this limit, the diffusive term vanishes and the non-diffusive term dominates. Vertical tracer mixing is significantly controlled by the correlation between the chemical equilibrium distribution and the vertical velocity distribution. If the correlation is positive, i.e., tracer is more abundant in the upwelling region than the downwelling region due to chemistry, and if the tracer is produced at the top, i.e., the vertical gradient of the mean tracer mixing ratio is positive, the derived $K_{zz}$ is negative (Eq. 16). A negative effective eddy diffusivity does not make sense physically as it suggests the tracer is mixed towards its source. The reason is simply because the global-mean vertical tracer transport is essentially non-diffusive in this case. 

For a given vertical gradient of global-mean tracer, $K_{zz}$ becomes larger if the circulation is stronger, the tracer chemical lifetime is longer, and the horizontal mixing timescale is longer. This trend is consistent with that in the case with uniform chemical equilibrium abundance. Interestingly, as the chemical timescale becomes longer, the diffusive term gets larger and the non-diffusive term becomes smaller, perhaps because the longer-lived tracers are more controlled by the dynamical transport so that the non-uniform chemical equilibrium has less effect. Thus the non-diffusive correction in Eq. (15) should work better for species with relatively long timescales. 

\subsection{Regime III: Long-lived Tracers}
If the tracer chemical lifetime is long and the tracer is almost inert, the material surfaces (tracer contours) are usually distorted significantly \footnote{If the tracer is absolutely inert with a very long lifetime, it will be completely homogenized over the globe by the horizontal transport. But in this study we do not investigate this type of real conservative, well-mixed tracers. The ``long-lived tracer" in our study stands for the species with a significantly long chemical lifetime so that the atmospheric dynamics greatly shapes its material surface, but the horizontal variation of the tracer distribution is still not small.}. The above discussion for short-lived tracers could be violated since the assumption (ii) in Section 2.1.1 might no longer be valid. It is expected that the 3D distribution of such a ``quasi-conservative'' tracer is significantly controlled the atmospheric dynamics. Due to the complicated dynamical transport behavior, no simple analytical theory of the global-mean vertical tracer transport exists, and thus the corresponding $K_{zz}$ is not generally known.

Although it is expected that there are some non-diffusive effects in this regime, the diffusive framework may still be useful. If we still apply the $K_{zz}$ theory in Section 2.3 to this regime and let $\tau_c\rightarrow \infty$, the non-diffusive contribution vanishes (Eq. 12). In this case, the influence of the chemical equilibrium tracer distribution is negligible and Eq. (15) is reduced to Eq. (13). One might expect $K_{zz}$ to approach its asymptotic value in Eq. (13): 
\begin{equation}
K_{zz}\approx\overline{w^2}\tau_{d}=\hat{w}L_v
\end{equation}
where we have introduced a vertical transport length scale $L_v=\hat{w}\tau_{d}$ and $\hat{w}$ is the root-mean-square of the vertical velocity $\hat{w}=(\overline{w^2})^{1/2}$. Here the vertical transport timescale is assumed as the horizontal tracer mixing timescale $\tau_{d}$ due to the continuity equation. Eq. (17) is in the similar form of that from the mixing length theory although there is no convective or small-scale mixing due to wave breaking in our theory. Using $L_v$, the effective eddy diffusivity (Eq. 13) for short-lived tracers with uniform chemical equilibrium (Section 2.3.1) can also be represented as:
 \begin{equation}
K_{zz}=\frac{\hat{w}L_v}{1+\tau_{d}\tau_{c}^{-1}}.
\end{equation}
The magnitude of the change in $K_{zz}$ depends on the ratio of the timescales between the dynamical and chemical processes: $\tau_{d}/\tau_{c}$. In the long-lived tracer regime where $\tau_c\rightarrow \infty$, the effective eddy diffusivity $K_{zz}$ approaches Eq. (17). 

The vertical transport length scale $L_v$ in our $K_{zz}$ theory cannot be arbitrarily chosen. It critically depends on the atmospheric dynamics, specifically the vertical velocity $\hat{w}$ and the horizontal dynamical timescale $\tau_d$. If the dynamical timescale $\tau_d$ for horizontal tracer mixing is equal to the global horizontal advection timescale $a/U$ where $a$ is approximately the planetary radius, through continuity $\tau_d$ should be approximately $H/\hat{w}$. Then $L_v$ would be equal to the pressure scale height $H$. If the horizontal mixing timescale is longer (or shorter) than the horizontal advection timescale, $L_v$ is then larger (or smaller) than $H$ by that same factor. Using the vertical velocity from general circulation models and simply assuming $L_v$ is $H$, some previous models (e.g., \citealt{lewis-etal-2010}, \citealt{moses-etal-2011}) estimated the eddy diffusivities on exoplanets based on Eq. (17). Those estimates were much larger than the eddy diffusivity derived based on 3D passive tracer simulations (\citealt{parmentier20133d}). 

There are two reasons. First, $K_{zz}$ estimated from Eq. (17) is the maximum eddy diffusivity one can obtain from Eq. (18). For short-lived tracers, the effective eddy diffusivity should be smaller than that from Eq. (17). Second, in the long-lived tracer regime, tracers are significantly controlled by atmospheric dynamics and the horizontal dynamical timescale $\tau_d$ might be different from the global horizontal advection timescale $a/U$. Thus the vertical characteristic transport length scale $L_v$ in Eq. (17) and (18) could be different from $H$. This has also been noted in the studies of tracer transport in the convective atmospheres. For example, \citet{smith1998estimation} investigated dynamical quenching of chemical tracers in convective atmospheres and found that vertical transport length scale in the traditional mixing length theory should depend on the chemical tracer equilibrium distribution as well as the chemical and dynamical timescales. Recent work by \citet{bordwell2018convective} explored the chemical tracer transport using non-rotating local convective box models. They found that using the chemical scale height as the vertical transport length scale, which is usually smaller than the scale height $H$, leads to a better prediction of the chemical quenching levels. 

We reiterate that the assumptions used to derive Eq. (17) likely break down in the regime of long-lived tracers, so it may be that the qualitative dependencies implied in Eq. (17) are not rigorously accurate when the tracers are long-lived. In our theory we have dropped the last term in the left hand side in Eq. (8), which might become important in the long-lived regime. As we will demonstrate in the numerical simulations later, this non-linear eddy term could potentially enhance or decrease the global tracer mixing efficiency. We also emphasize that, although we adopt the diffusive framework here in the long-lived tracer regime, the tracer transport in this regime may not always behave diffusively. As we will also show later in the numerical simulations, the diffusive assumption could break down in some cases when the tracer material surface is distorted significantly and non-diffusive effects are substantial.

In sum, in this section we have developed an approximate analytical theory for the 1D global-mean tracer transport in a 3D atmosphere. We demarcated three atmospheric regimes in terms of the tracer chemical lifetime and horizontal tracer distribution under chemical equilibrium. The underlying physical mechanisms governing the global-mean vertical tracer transport in the three regimes are different. The traditional chemical-diffusion assumption is mostly valid in the first regime but could be violated in the second and third regimes. We predicted the analytical expression of the 1D effective eddy diffusivity $K_{zz}$ for the global-mean vertical tracer transport. $K_{zz}$ depends on both atmospheric dynamics and tracer chemistry. Crudely speaking, if the atmospheric dynamics is fixed, $K_{zz}$ roughly scales with the tracer chemical lifetime $K_{zz}\propto\tau_c$ (Eq. 13) when the tracer lifetime is short (regime I) and approaches to a constant value when the tracer lifetime is long (regime III, Eq. 17). We also found that in the short-lived tracer regime (regime I), $K_{zz}$ roughly scales with the the vertical velocity square $K_{zz}\propto\hat{w}^2$ (Eq. 13), but in the long-lived tracer regime (regime III), it scales with root-mean-square of the vertical velocity $K_{zz}\propto\hat{w}$ (Eq. 17). Next, we will perform a series of numerical experiments to quantitatively verify our theoretical arguments. We will primarily focus on stratified atmospheres here.

\section{2D Simulations on Fast-rotating Planets}

\begin{deluxetable*}{ccccccc}[bp]
\centering
\tablecolumns{4} 
\tablewidth{0pt}  
\tablecaption{2D Simulation cases in this study.}
\tablehead{\colhead {Experiment}			&
\colhead {Streamfunction}  &
\colhead {$K_{yy2D}$ ($\mathrm{m^2~s^{-1}}$ )} &
\colhead {Tracer source} & 
\colhead {Latitudinal distribution of $\chi_0$} 
		}
\startdata
I    & $\psi_A$       & $10$     & Deep     & Uniform  &\\
II    & $\psi_B$      & $10$     & Deep     & Uniform  &\\
III    & $\psi_A$      & $10$    & Top     & Uniform  &\\
IV     & $\psi_A$     & $10^6$     & Deep     & Uniform  & \\
V     & $\psi_A$     & $10$     & Top     & Non-uniform  &
\enddata
\end{deluxetable*}

Now we consider numerical simulations of tracer transport on a fast-rotating planet on which the tracer is uniformly distributed with longitude (but not necessarily with latitude). Most of the planetary atmospheres in the solar system are close to this situation. This is essentially a 2D tracer transport problem that can be studied using a 2D zonally symmetric model. In our numerical simulations, we will only study passive tracers, i.e., no radiative feedback from the tracer to the dynamics. 

2D zonal-mean tracer transport has been extensively discussed in the Earth literature (e.g., \citealt{holton1986dynamically}). But till now there has not been a thorough and specific study on the global-mean eddy diffusivity using 2D numerical simulations with chemical tracers. On a fast-rotating planet, when averaged zonally, the meridional circulation that transports the tracer (like the one shown in Fig. 2) should be considered as the ``Lagrangian mean circulation'', which can be approximated by a Transformed Eulerian Mean (TEM) circulation (\citealt{andrews1976planetary}), and called the ``residual mean circulation''. Another formalism of the zonal-mean tracer transport introduced by \citet{plumb1987zonally} used the ``effective transport velocity". The difference between the two circulation formalisms is usually small in Earth's stratosphere and vanishes if the waves are linear, steady, and adiabatic (\citealt{AHL1987}). 

In both frameworks, the zonal-mean eddy tracer fluxes can be parameterized as diffusive fluxes using a symmetric ``diffusion tensor" with four parameters (diffusivities): $K_{yy2D}$, $K_{zz2D}$, $K_{yz2D}$ and $K_{zy2D}$ (\citealt{AHL1987}). $K_{yz2D}$ and $K_{zy2D}$ are negligible in isentropic coordinates (\citealt{tung1982two}). As here we mainly focus on the stratified atmosphere where the inclination between the isobars and isentropes is usually small, we can ignore $K_{yz2D}$ and $K_{zy2D}$ in 2D chemical tracer transport simulations in this study (e.g., \citealt{garcia1983numerical}, \citealt{shia1989sensitivity}). 

A 3D model would naturally produce small-scale eddies that would cause mixing, and if the eddies were fully resolved, no parameterization of diffusive fluxes would be needed in the tracer transport. But in our 2D framework that ignores any role for such eddies, even for a planet with a great degree of zonal symmetry at large scales (like Earth or Jupiter), we have to parameterize the horizontal and vertical diffusivities $K_{yy2D}$ and $K_{zz2D}$ as a separately included process. In principle $K_{yy2D}$ and $K_{zz2D}$ can be estimated from the Eliassen-Palm flux (\citealt{AHL1987}) in a 3D model. But for our purpose of studying the response of the chemical tracer distribution to the dynamics and the global-mean vertical tracer mixing, we just prescribe the diffusivities and the circulation. 

\begin{figure*}
\includegraphics[width=0.95\textwidth]{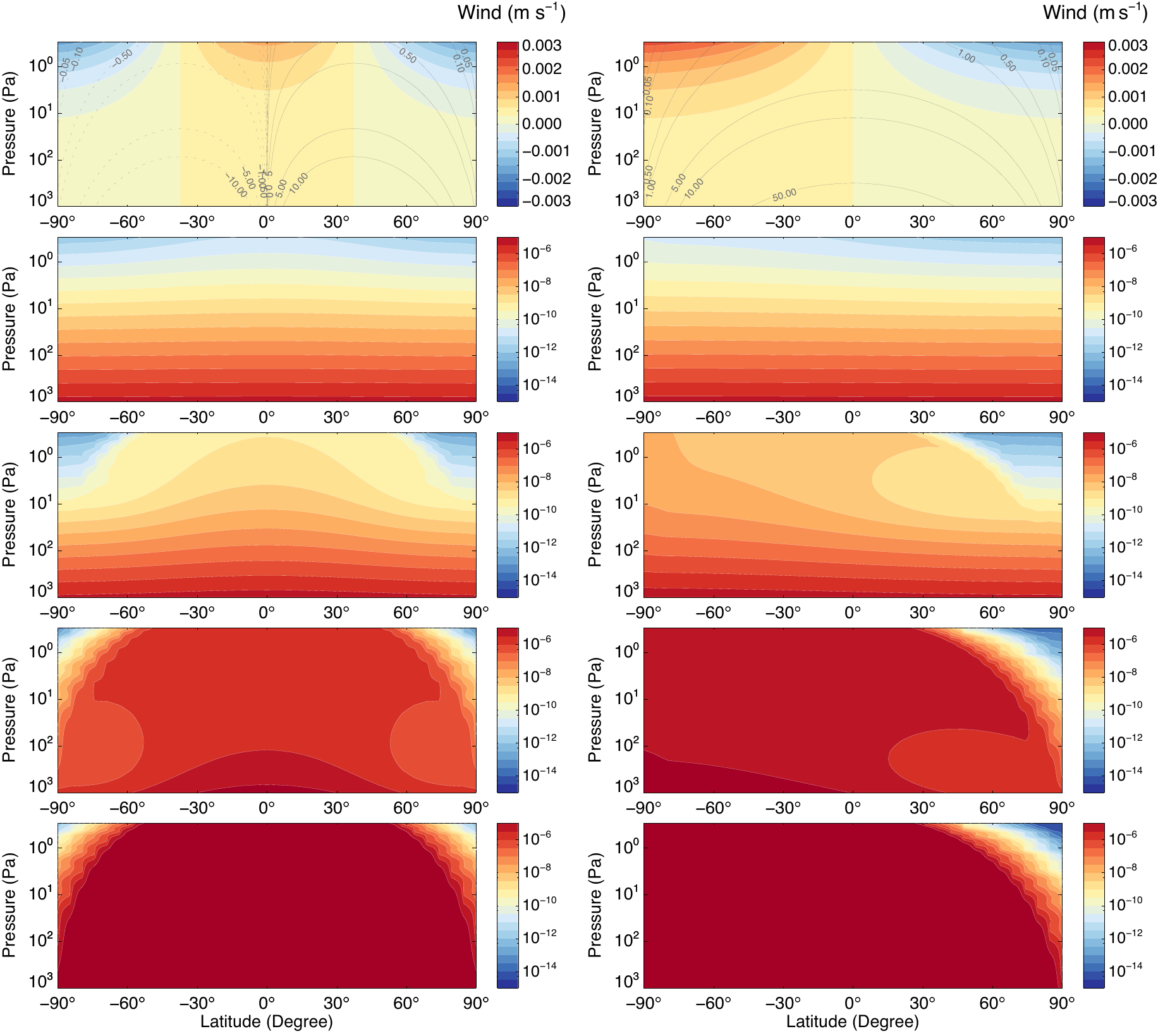} 
 \centering\caption{Latitude-pressure maps of experiments I (left) and II (right). First row: vertical winds (color) and mass streamfunctions (contours, in units of $10^{13}~\mathrm{Kg~s^{-1}}$). Starting from the second row we show volume mixing ratio maps of tracers with chemical timescales of $10^7$ s, $10^8$ s, $10^9$ s and $10^{10}$ s from top to the bottom, respectively.} 
\end{figure*}

In this study, we prescribe the temperature structure and circulation pattern and hold them constant with time. The governing equation of a 2D chemical transport system using the coordinates of log-pressure and latitude, can be written as (\citealt{shia1989sensitivity}, \citealt{shia1990two}, \citealt{zhang2013jovian}): 
\begin{equation}
\begin{split}
\frac{\partial \chi}{\partial t}+v^*\frac{\partial \chi}{\partial y}+w^*\frac{\partial \chi}{\partial z}-\frac{1}{\cos\phi}\frac{\partial}{\partial y}(\cos\phi K_{yy2D}\frac{\partial\chi}{\partial y}) \\
-e^{z/H}\frac{\partial}{\partial z}(e^{-z/H}K_{zz2D}\frac{\partial\chi}{\partial z})=S
\end{split}
\end{equation}
where $\phi=y/a$ is latitude and $a$ is the planetary radius. $H$ is the pressure scale height. The chemical source and sink term $S$ follows the linear relaxation scheme of Eq. (9). 

Residual circulation velocities are $v^*$ and $w^*$ in the meridional and vertical directions, respectively. For a 2D circulation pattern, we can introduce a mass streamfunction $\psi$ such that:
 \begin{subequations}
\begin{align}
v^*&=-\frac{1}{2\pi a\rho_0\cos\phi}e^{z/H}\frac{\partial}{\partial z}(e^{-z/H}\psi)
\\
w^*&=\frac{1}{2\pi a\rho_0\cos\phi}\frac{\partial \psi}{\partial y}
\end{align}
\end{subequations}
where $\rho_0$ is the reference density of the atmosphere at log-pressure $z=0$. With prescribed distributions of $\psi$, $K_{yy2D}$ and $K_{zz2D}$, we solve the governing equations using the Caltech/JPL 2D kinetics model (for numerics, refer to \citealt{shia1990two}). We use the Prather scheme for the 2D tracer advection (\citealt{prather1986numerical}). The model has been rigorously tested against multiple exact solutions under various conditions (\citealt{shia1990two}, \citealt{zhang2013jovian}). 

In this work, we adopted two different meridional circulation patterns (see Section 4 in \citealt{zhang2013jovian}). The circulation $\psi_A$ is an equator-to-pole pattern and circulation $\psi_B$ is pole-to-pole, corresponding to planets with low obliquity and high obliquity, respectively:
 \begin{subequations}
\begin{align}
\psi_A&=2\pi a^2\rho_0w_0e^{\eta z/H}\sin\phi\cos^2\phi\\
\psi_B&=2\pi a^2\rho_0w_0e^{\eta z/H}\cos^2\phi.
\end{align}
\end{subequations}
But we do not investigate the seasonal change of the circulation pattern in this study. The above circulation patterns are assumed steady with time in our simulations.

The mass streamfunctions and vertical velocities of the two circulation patterns are shown in Fig. 4. For both circulations, the area-weighted global-mean vertical velocity scale is about $\gamma w_0e^{\eta z/H}$ where $\gamma$ is an order-unity pre-factor originating from the global average.\footnote{$\gamma$ is $\mathrm{2/\sqrt{5}}$ for $\psi_A$ and $\mathrm{2/\sqrt{3}}$ for $\psi_B$ from the global integrations of Eq. (21a) and (21b), respectively. Thus the effective transport is a bit stronger using the circulation pattern B.} 

In this 2D chemical-advective-diffusive system, both the meridional advection and horizontal eddy diffusion contribute to the horizontal tracer mixing, and both the vertical advection and vertical eddy diffusion contribute to the vertical tracer transport. The 1D effective eddy diffusivity $K_{zz}$ in this system can be analytically predicted following the procedure introduced in Section 2. But since here we have to treat explicitly the eddy diffusion terms $K_{yy2D}$ and $K_{zz2D}$ originating from zonal-mean 2D dynamics, we provided a detailed derivation of $K_{zz}$ in this 2D system in the Appendix. We show that the vertical eddy diffusion by $K_{zz2D}$ in Eq. (19) can be treated as an additive term in the global-mean effective eddy diffusivity $K_{zz}$. 

In the 2D framework, the $K_{zz}$ for the situation of non-uniform chemical equilibrium mixing ratio $\chi_{0}$ can be expressed as (See Appendix for more details): 
  \begin{equation}
  \begin{aligned}
K_{zz}&\approx K_{zz2D}+\frac{\gamma^2w_0^2e^{2\eta z/H}}{K_{yy2D} a^{-2}+\gamma w_0e^{\eta z/H} H^{-1}+\tau_{c}^{-1}}\\
&-\frac{\gamma w_0e^{\eta z/H}\Delta{\chi_0^{\prime}}}{1+\tau_{c}K_{yy2D} a^{-2}+\tau_{c}\gamma w_0e^{\eta z/H} H^{-1}}(\frac{\partial\overline{\chi_0}}{\partial z})^{-1}
 \end{aligned}
\end{equation}
where $\overline{\chi_0}$ is the global-mean of the non-uniform chemical equilibrium mixing ratio $\chi_{0}$. $\Delta{\chi_0^{\prime}}$ is the root-mean-square of the deviation $\chi_0^{\prime}$ over the globe. For a cosine function of $\chi_0$, $\Delta{\chi_0^{\prime}} \approx 0.28~\overline{\chi_0}$. Here we have also assumed the horizontal transport length scale $L_h\sim a$ and vertical transport length scale $L_v\sim H$. If $\chi_{0}$ is uniform, $K_{zz}$ can be reduced to:
\begin{equation}
K_{zz}\approx K_{zz2D}+\frac{\gamma^2 w_0^2e^{2\eta z/H}}{K_{yy2D} a^{-2}+\gamma w_0e^{\eta z/H} H^{-1}+\tau_{c}^{-1}}.
\end{equation}

In the second term on the right hand side, the three terms in the denominator correspond to horizontal tracer diffusion due to parameterized effects of eddies and waves, horizontal tracer advection by zonal-mean flow and tracer chemistry, respectively. 

We use Jupiter's parameters in these 2D simulations but with an isothermal atmosphere of 150 K from $3000$ Pa to about 0.2 Pa for simplicity. For the circulation patterns, we adopt $w_0=10^{-5}~\mathrm{m~s^{-1}}$ and $\eta=0.5$. Given a pressure scale height $H$ of about 25 km, the vertical advection timescale changes from $10^9$ s at the bottom to about $10^7$ s at the top. The vertical profile of the advection timescale is similar to the empirical vertical mixing timescale derived from $H^2/K_z$ where $K_z$ is the empirical eddy diffusivity in the 1D stratospheric chemical model on Jupiter (\citealt{moses2005photochemistry}, also see Fig. 1). The horizontal advection timescale in our model is about $10^8$ s at 100 Pa, comparable to the vertical circulation timescale due to the continuity constraint. This horizontal transport timescale is also similar to that inferred from Voyager and Cassini observations (\citealt{nixon2010abundances}, \citealt{zhang2013radiative}). 

We designed five 2D experiments (Table 1) and simulated 9 tracers in each experiment. In all cases, we set the vertical diffusion coefficients $K_{zz2D}$ zero because here we mainly investigate the influence of the large-scale overturning circulation, horizontal mixing and chemical source/sink on the global-mean vertical tracer transport in this study. According to Eq. (23), the influence of $K_{zz2D}$ in the 2D simulation is trivial because it can just be treated as an additive term in $K_{zz}$. We also tested different horizontal diffusional coefficients $K_{yy2D}$. 

Experiments I-IV are simulations with chemical tracers with uniform chemical equilibrium mixing ratios. Experiment V focuses on the chemical tracer transport with non-uniform chemical equilibrium. Experiment I (Fig. 4) is the standard case with an equator-to-pole streamfunction $\psi_A$ and a constant horizontal eddy diffusivity $K_{yy2D} = 10~\mathrm{m^2~s^{-1}}$. This is a ``deep source'' case in which tracers are advected upward from the deep atmosphere at $3000$ Pa. The vertical profile of equilibrium tracer mixing ratio $\chi_{eq}$ is assumed to be a power-law function of pressure $\chi_{eq}(p)=10^{-5}(p/p_0)^{1.7}$ where $p_0=3000$ Pa is the pressure at the bottom boundary. $\chi_{eq}$ starts from $10^{-5}$ at bottom and decreases towards $10^{-12}$ at the top of the atmosphere. A linear chemical scheme (Eq. 9) is applied to each tracer with a relaxation timescale $\tau_c$ that varies from $10^7$ to $10^{11}$ s. The chemical timescale of the $i$th tracer is assumed as $\tau_c=10^{6.5+0.5i}$ s. 

\begin{figure}[t]
  \centering \includegraphics[width=0.48\textwidth]{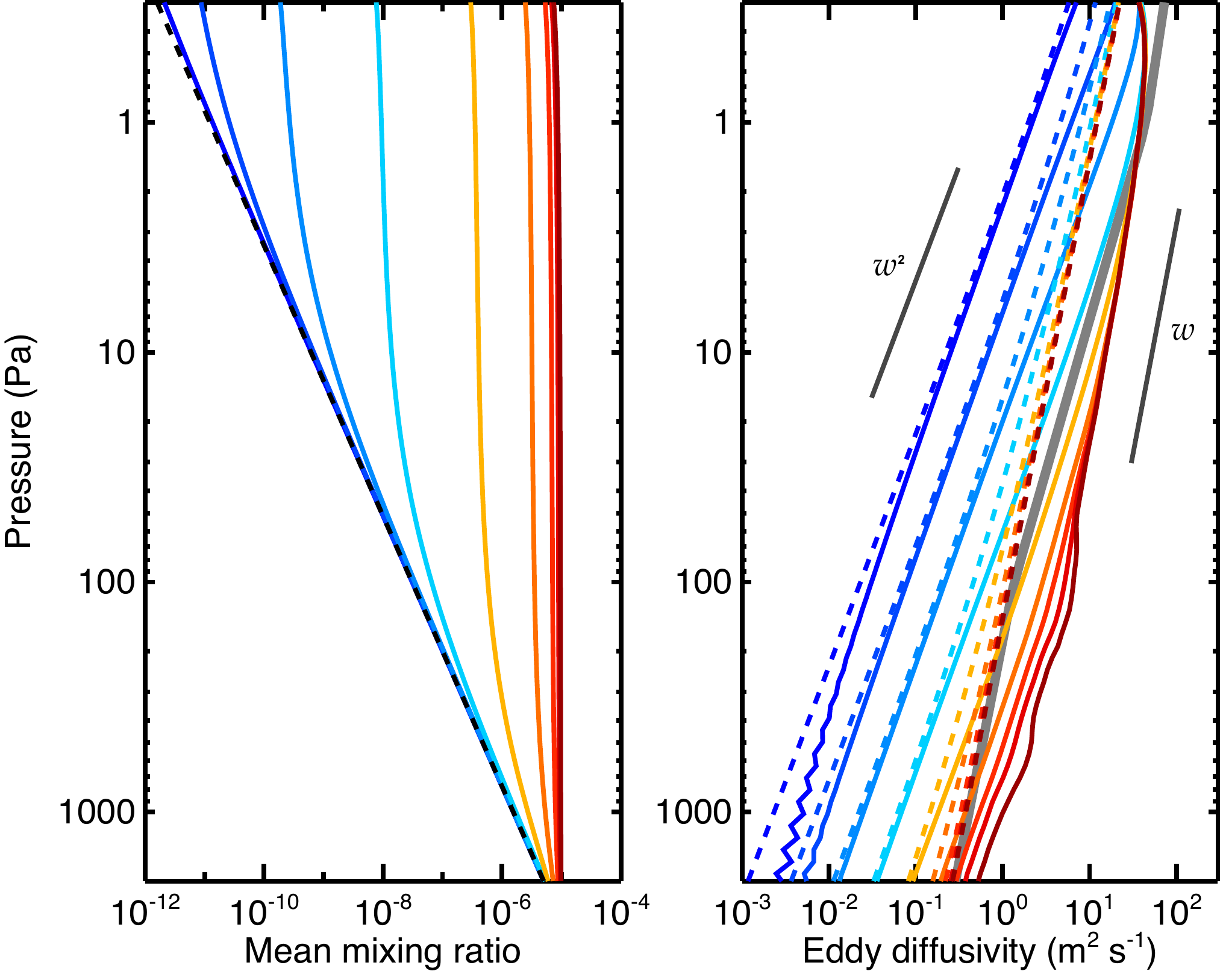} 
  \caption{Vertical profiles of the global-mean volume mixing ratio (left) and derived effective eddy diffusivity (right) from Experiment I. Different colors from cold (blue) to warm (red) represent tracers with different chemical timescales from short to long, ranging from $10^7$ s to $10^{11}$ s. The prescribed equilibrium tracer mixing ratio profile is shown in the dashed line in the left panel, which is nearly on top of the mixing ratio profile of the very short-lived tracer (dark blue solid line, see the upper left corner of the left panel). The predicted eddy diffusivity profiles based on Eq. (23) are shown in dashed in the right panel. The solid lines are derived from the simulations. The thick gray line indicates the empirical eddy diffusivity profile used in current photochemical models in Jupiter's stratosphere (\citealt{moses2005photochemistry}, \citealt{moses2017dust}).} 
\end{figure}

\begin{figure}[t]
  \centering \includegraphics[width=0.48\textwidth]{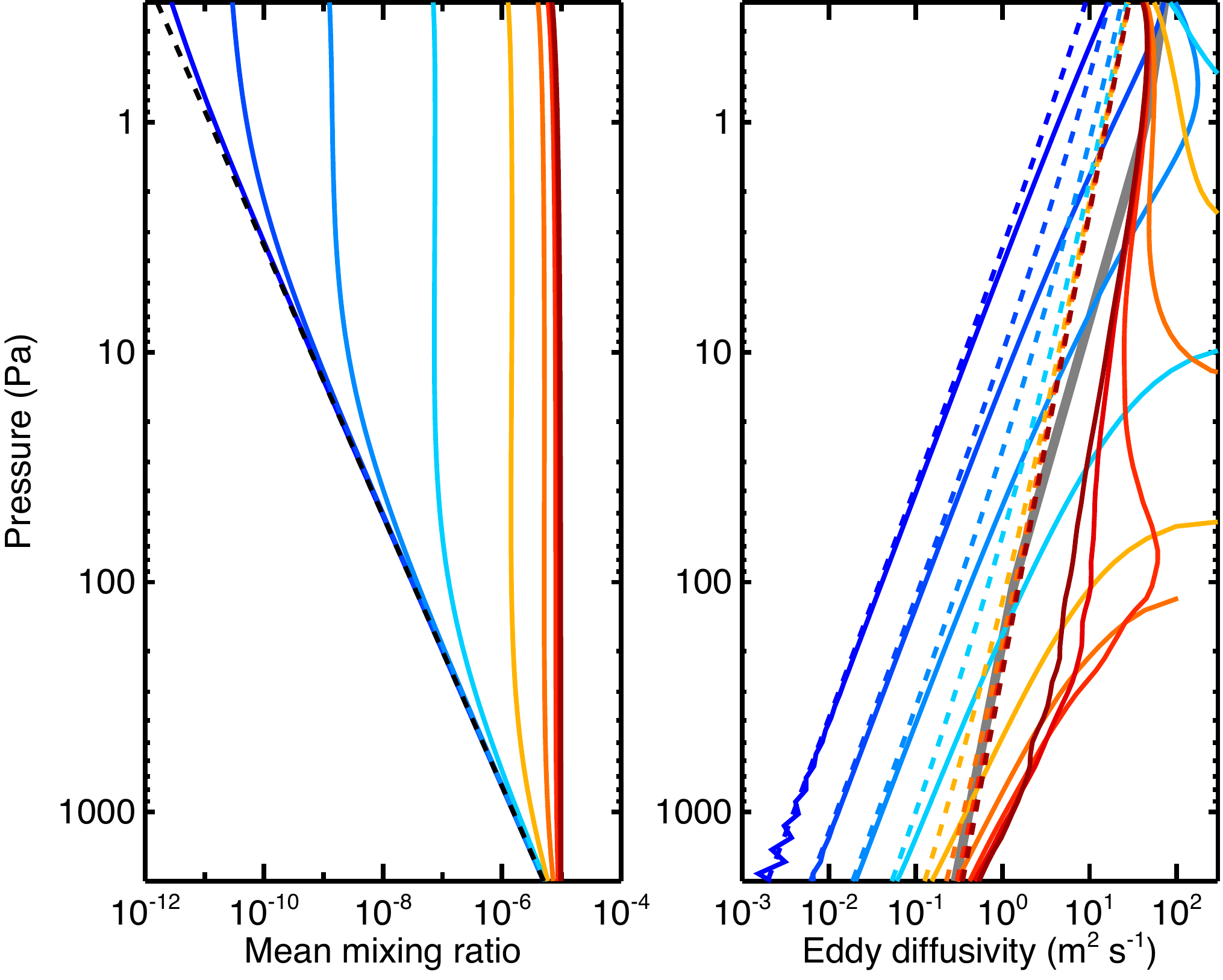} 
  \caption{Same as Fig. 5 but from Experiment II. Note that some curves (e.g., the orange line) break apart in the long-live tracer regime. This is because the material surface is distorted so large that the derived $K_{zz}$ is negative in the middle part.} 
\end{figure}

Experiment II is same as Experiment I but with a different circulation pattern $\psi_B$. Experiment III is also same as Experiment I but with a ``top chemical source''. The equilibrium tracer mixing ratio, while still uniform across the latitude, is large at top ($10^{-5}$) and small at the bottom ($10^{-12}$). In Experiments I to III, the horizontal eddy diffusion timescale is $a^2/K_{yy2D}\sim 5\times10^{14}$ s. The diffusive transport is much less efficient compared with the horizontal advection and can be neglected (timescale of $10^7-10^9$ s). We designed Experiment IV using the streamfunction $\psi_A$ but horizontal eddy diffusivity $K_{yy2D}$ enhanced to $10^6~\mathrm{m^2~s^{-1}}$. In this case, the diffusive timescale is about $5\times10^9$ s, comparable to the circulation timescale (but the $K_{zz2D}$ is still zero). In this setup we can test the influence of the horizontal eddy mixing to the global-mean vertical tracer transport. Experiment V is similar to Experiment III with a ``top chemical source'' but the tracer chemical equilibrium mixing ratios are not uniformly distributed with latitude. The non-uniform chemical equilibrium mixing ratio is assumed to be a cosine function of the latitude: $\chi_0(p, \phi)=\chi_{eq}(p)\cos\phi$. This setup is closer to the realistic photochemical production in planetary atmospheres where the photochemical photon flux changes with latitude. 

We discretized the atmosphere into 35 latitudes in the horizontal dimension, corresponding to  $5^\circ$ per grid cell. Vertically, the log-pressure grid is evenly spaced in 80 layers from $3000$ Pa to about 0.2 Pa. The time step in the simulations is $10^5$ second. We ran the simulations for about $10^{13}$ s to ensure the spatial distributions of the tracers have reached the steady state. The tracer abundances were averaged over the last $10^9$ s for analysis. We tested the model with different vertical and horizontal resolutions to confirm that the simulation results are robust.

\begin{figure}[t]
  \centering \includegraphics[width=0.48\textwidth]{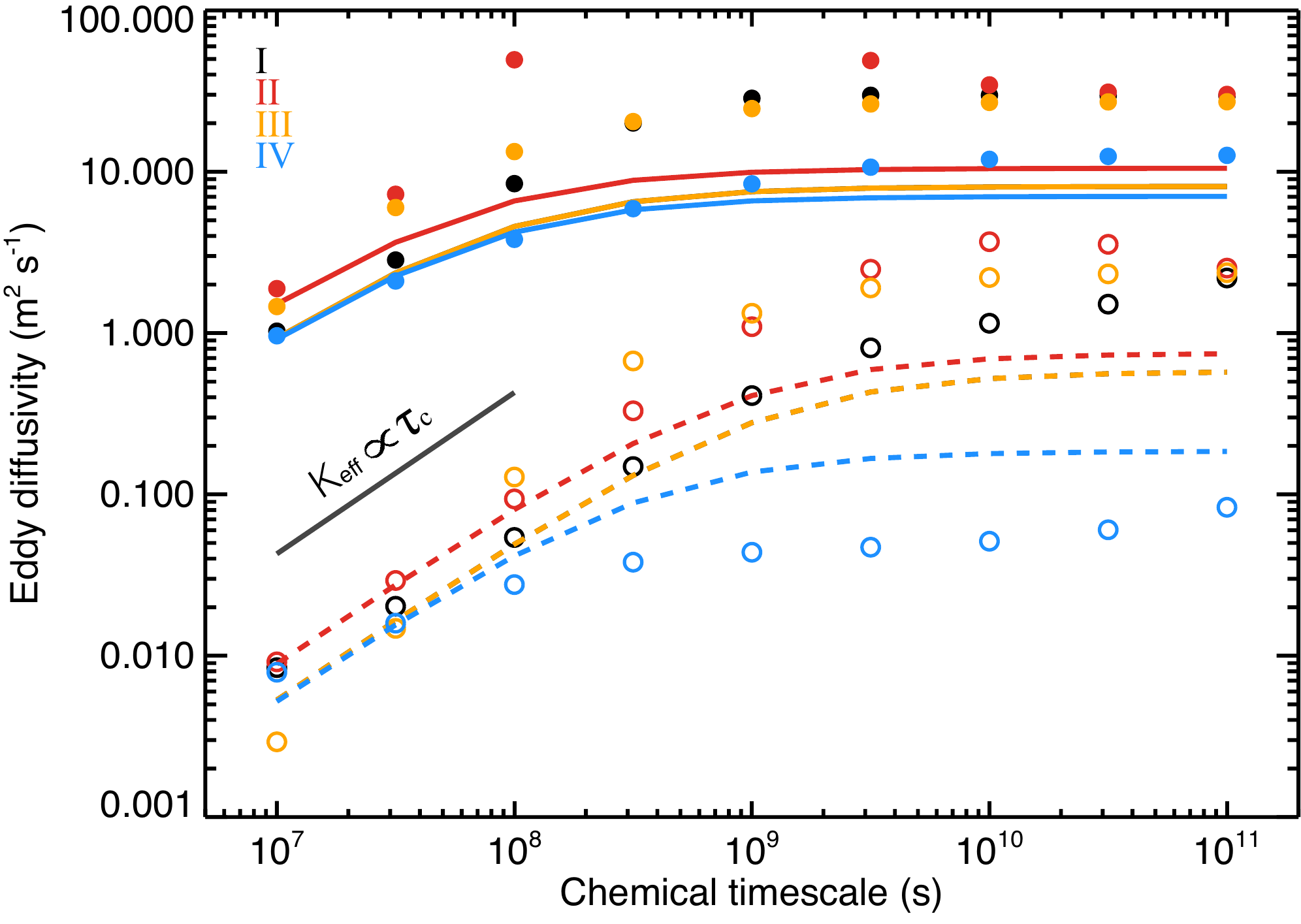} 
  \caption{$K_{zz}$ as a function of tracer chemical timescale for all four experiments at two typical pressure levels, 80 Pa (filled circles) and 200 Pa (open circles). The solid and dashed lines are the predictions from Eq. (23).} 
\end{figure}

\begin{figure}[t]
  \centering \includegraphics[width=0.48\textwidth]{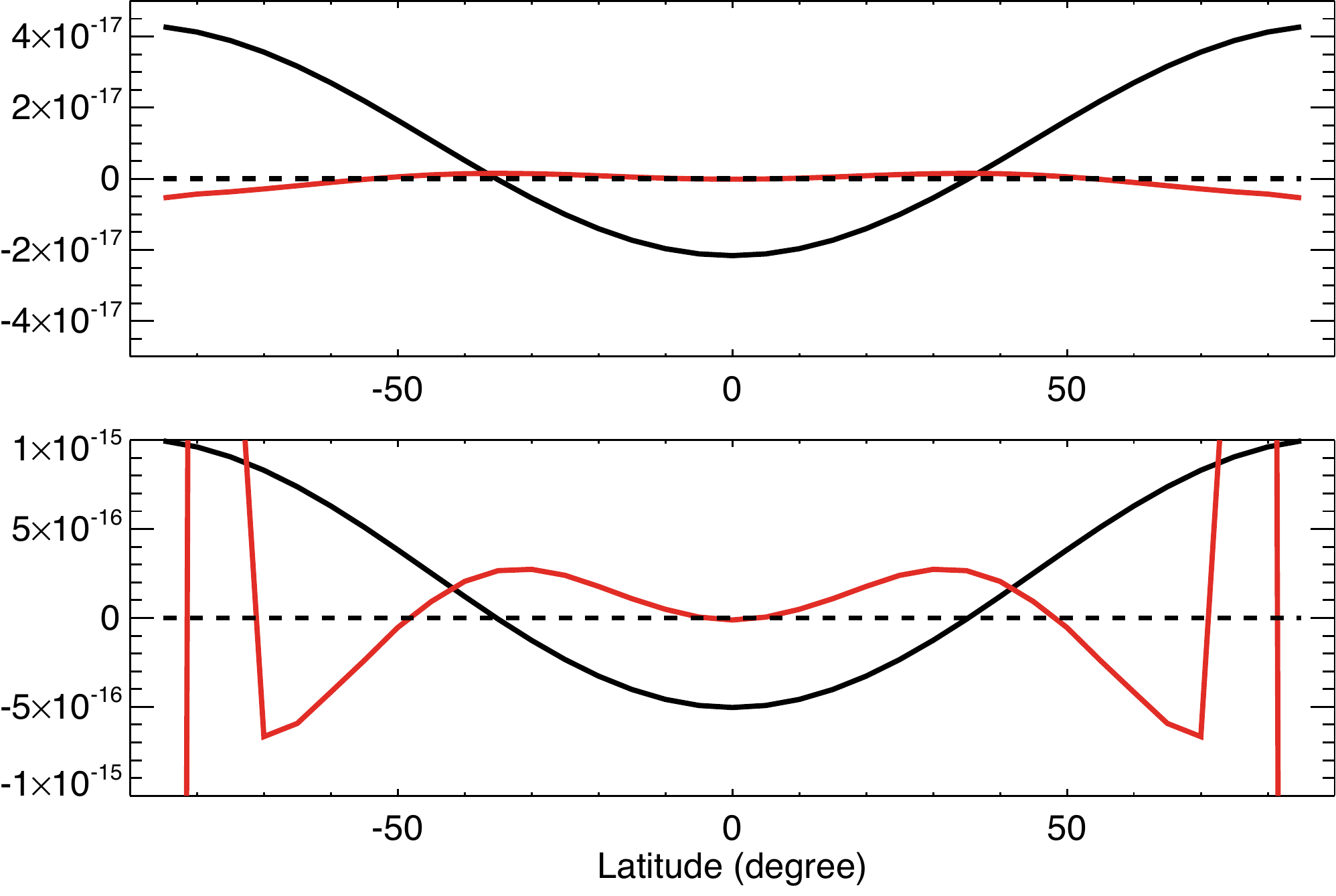} 
  \caption{Important tracer tendency terms in Eq. (8) for a short-lived tracer (upper panel, $\tau_{c}=10^{7}$ s) and a long-lived tracer (lower panel, $\tau_{c}=10^{10}$ s) in Experiments I. The black lines represent the term $w\frac{\partial\overline{\chi}}{\partial z}$ and the red are $e^z\frac{\partial}{\partial z}[e^{-z}(w\chi^\prime-\overline{w\chi^\prime})]$.} 
\end{figure}

\begin{figure*}
\includegraphics[width=0.95\textwidth]{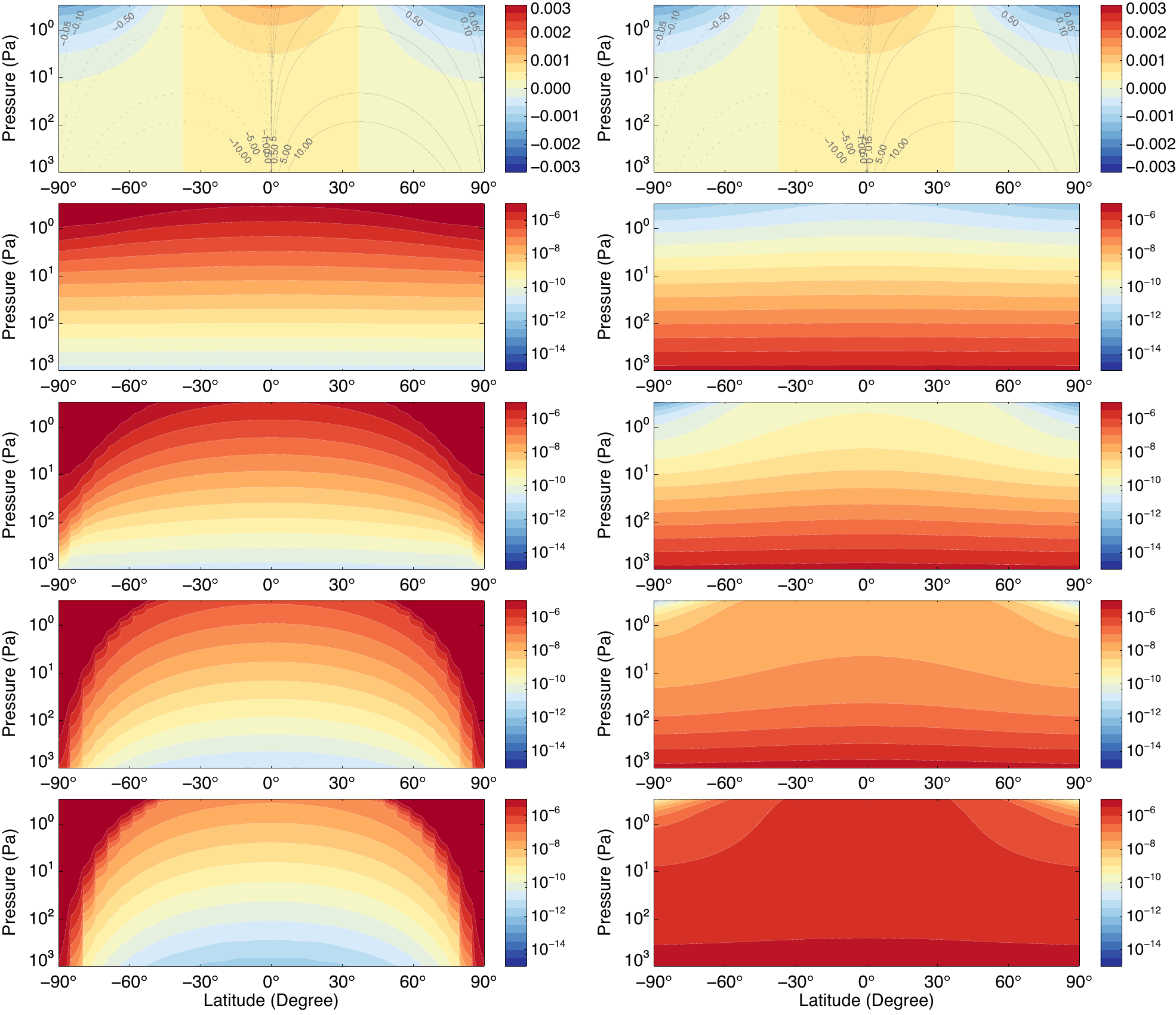} 
 \centering\caption{Same as Fig. 4 but for Experiments III (left) and IV (right).} 
\end{figure*}

\subsection{Results: Simulations with Uniform $\chi_0$ (Experiments I-IV)}

The steady state results in Experiments I and II are shown in Fig. 4. Although the final tracer distributions are different between the two experiments due to different circulation patterns, some common behaviors exist. The short-lived tracers are uniformly distributed across latitude. As the tracer chemical lifetime increases, the effect of the circulation becomes stronger. In the upwelling region, tracers with higher mixing ratios are transported upward from their deep source; while in the downwelling region, tracers with lower mixing ratios are transported from the upper atmosphere. As a result, the tracer mixing ratio is higher in the upwelling region and lower in the downwelling region on an isobar. The final latitudinal variations of the short-lived tracers follow the pattern of the vertical velocity (Fig. 4 and Eq. 14). However, if the tracer chemical timescale is very long (Fig. 4, bottom row), the tracers tend to be homogenized by the circulations. 

We numerically derive $K_{zz}$ based on the simulation results and the flux-gradient relationship (Eq. 6) and compare with the analytical prediction (Eq. 23). First, we averaged the tracer distributions with latitude in an area-weighted-mean fashion. The vertical profiles of the global-mean tracer mixing ratios are shown in Fig. 5 for Experiment I and Fig. 6 for Experiment II, respectively. The vertical profiles of short-lived tracers are close to the chemical equilibrium profile but that of the long-lived tracers are almost well mixed and ``quench" to the lower atmosphere values because the global circulations efficiently smooth out their vertical gradients. We then derived the 1D effective eddy diffusivity $K_{zz}$ by equating the eddy diffusive flux to the global-mean net vertical flux of the tracers (Eq. 6, Fig. 5 and Fig. 6). Our analytical prediction based on Eq. (23) matches the numerical results well. 

$K_{zz}$ increases from the bottom towards the top of the atmosphere because the vertical velocity is larger and transport is stronger in the upper atmosphere. Our theory predicts that $K_{zz}$ roughly scales with the square of the vertical velocity $K_{zz}\propto\hat{w}^2$ in the short-lived tracer regime (regime I, Eq. 13) and $K_{zz}\propto\hat{w}$ in the long-lived tracer regime (regime III, Eq. 17). The vertical profiles of the $K_{zz}$ follow the scaling very well (Fig. 5). The theory also predicted that $K_{zz}$ should be a strong function of chemical timescale (Eq. 13). This is also confirmed by the numerical simulations. $K_{zz}$ can increase by more than a factor of 1000 from short-lived tracers to long-lived tracers (Fig. 5 and Fig. 6). This increasing trend is well predicted by our analytical theory Eq. (23). Longer-lived tracers are more dynamically controlled than chemically controlled and therefore have better correlation with the circulation pattern, leading to a larger global-mean effective vertical transport. As we pointed out in Section 2, if the atmospheric dynamics is fixed (i.e., passive tracer transport), $K_{zz}$ scales with the tracer chemical lifetime $K_{zz}\propto\tau_c$ (Eq. 13) when the tracer lifetime is short (regime I) and approaches to a constant value when the tracer lifetime is long (regime III, Eq. 17).  Therefore, for tracers with very long chemical timescale, $K_{zz}$ is insensitive to the chemical timescale (also see Eq. 23). The dependence of $K_{zz}$ on $\tau_c$ is clearly seen in Fig. 7.

In Experiments I and II, horizontal diffusive transport via $K_{yy2D}$ is small compared with the wind advection. When $\tau_c\rightarrow \infty$, the effective eddy diffusivity (Eq. 21) approaches $K_{zz}\approx Hw_0e^{\eta z/H}$. The analytical $K_{zz}$ converges to a profile in Fig. 5, but does not match the numerical $K_{zz}$ exactly. There may be greater disparity in Experiment II (Fig. 6) where the numerical $K_{zz}$ exhibit wavy fluctuations. In the long-lived tracer regime, the derived diffusivity decreases with chemical timescale in some pressure ranges and even becomes negative at some pressure levels. This reveals a drawback in our theory in Section 2. As noted in Section 2.2.2, when the chemical lifetime is too long compared with the circulation timescale, the material surfaces of tracers are distorted significantly. The expression of $K_{zz}$ might no longer be valid because the last term (eddy term) in the left hand side of Eq. (8), $e^z\frac{\partial}{\partial z}[e^{-z}(w\chi^\prime-\overline{w\chi^\prime})]$, is comparable to or even larger than the third term (mean tracer term) $w\frac{\partial\overline{\chi}}{\partial z}$ and thus it cannot be neglected. As illustrated in Fig. 8, for a short-lived tracer with $\tau_{c}=10^7$ s, the eddy term is much smaller than the mean tracer term across the latitude and Eq. (23) is a good prediction of $K_{zz}$. But for a long-lived tracer with $\tau_{c}=10^{10}$ s, the eddy term is comparable to the mean tracer term. The eddy transport is complicated in this regime and may lead to some wavy features in the numerical $K_{zz}$ in Figs. 4 and 5, which our current analytical theory is unable to capture although the theoretical prediction is still within a factor of 2-5 in Experiment I and can be as large as a factor of 10 in Experiment II. 

Note that the theoretical prediction in the long-lived tracer regime generally underestimates the eddy mixing from the numerical simulations. In this regime, the theoretical $K_{zz}\approx \hat{w}L_v$ (Eq. 17). Here we have assumed that the horizontal dynamical timescale $\tau_d$ is equal to the global advection timescale $a/U$ and thus the vertical transport length scale $L_v\approx H$ through continuity. The discrepancy between the analytical and numerical eddy diffusivities implies that the actual horizontal dynamical timescale is larger than $a/U$ but the mechanism is not clear. A detailed future investigation is needed. 

In Experiment II (Fig. 4, third row, $\tau_c=10^9$ s case), the overturning circulation transports the high-concentration tracers from the deep atmosphere in the southern hemisphere all the way to the top of the northern hemisphere. These tracers are then mixed downward in the northern hemisphere above 100 Pa. The global-mean mixing ratio profile of these tracers also shows a local shallow minimum at around 100 Pa (yellow line in Fig. 6). This type of local minima is also commonly found in global-mean tracer mixing ratio profiles in the simulations of convective atmospheres (see Fig. 6 in \citealt{bordwell2018convective}). At the pressure levels right above this local minimum, because the tracer mixing ratio increases with altitude but the global-mean tracer transport is still upward, the global-mean tracer flux is against the local vertical gradient of the tracer. As a result, the predicted eddy diffusivity $K_{zz}$ is negative above 100 Pa (Fig. 6). As we have discussed in Section 2.3.2, this type of ``negative eddy diffusivity" phenomenon probably indicates that the global-mean vertical tracer transport in this case does not behave diffusively. Therefore, the diffusive assumption could break down in the long-lived tracer regime when the material surfaces are significantly distorted.  

When a uniform chemical source is located at the top (Experiment III), the horizontal distribution of the tracer is anti-correlated with the vertical velocity distribution (Fig. 9), i.e., the tracer abundance is higher in the downwelling region and lower in the upwelling region, as measured on isobars. This does not alter our theory of global-mean vertical mixing in Section 2.2.1. As before, the derived effective eddy diffusivities increase with tracer chemical timescale and approaches constant in the long-lived tracer regime (Fig. 7).  If we enhance the horizontal diffusion via $K_{yy2D}$ (Experiment IV), the diffusion tends to smooth out the horizontal gradient of tracer, leading to a weaker correlation between the tracer distribution and the vertical velocity. The latitudinal distributions of tracers (Fig. 9) are flatter in this case compared than that in Experiment I. The $K_{zz}$ is thus smaller (Fig. 7) but still increases with the chemical timescale. The trend is consistent with other experiments and our theory. As seen in Fig. 7, the decrease of $K_{zz}$ with a larger $K_{yy2D}$ can also be predicted in Eq. (23).

\subsection{Results: Simulations with Non-uniform $\chi_0$ (Experiment V)}
\begin{figure}[t]
\includegraphics[width=0.48\textwidth]{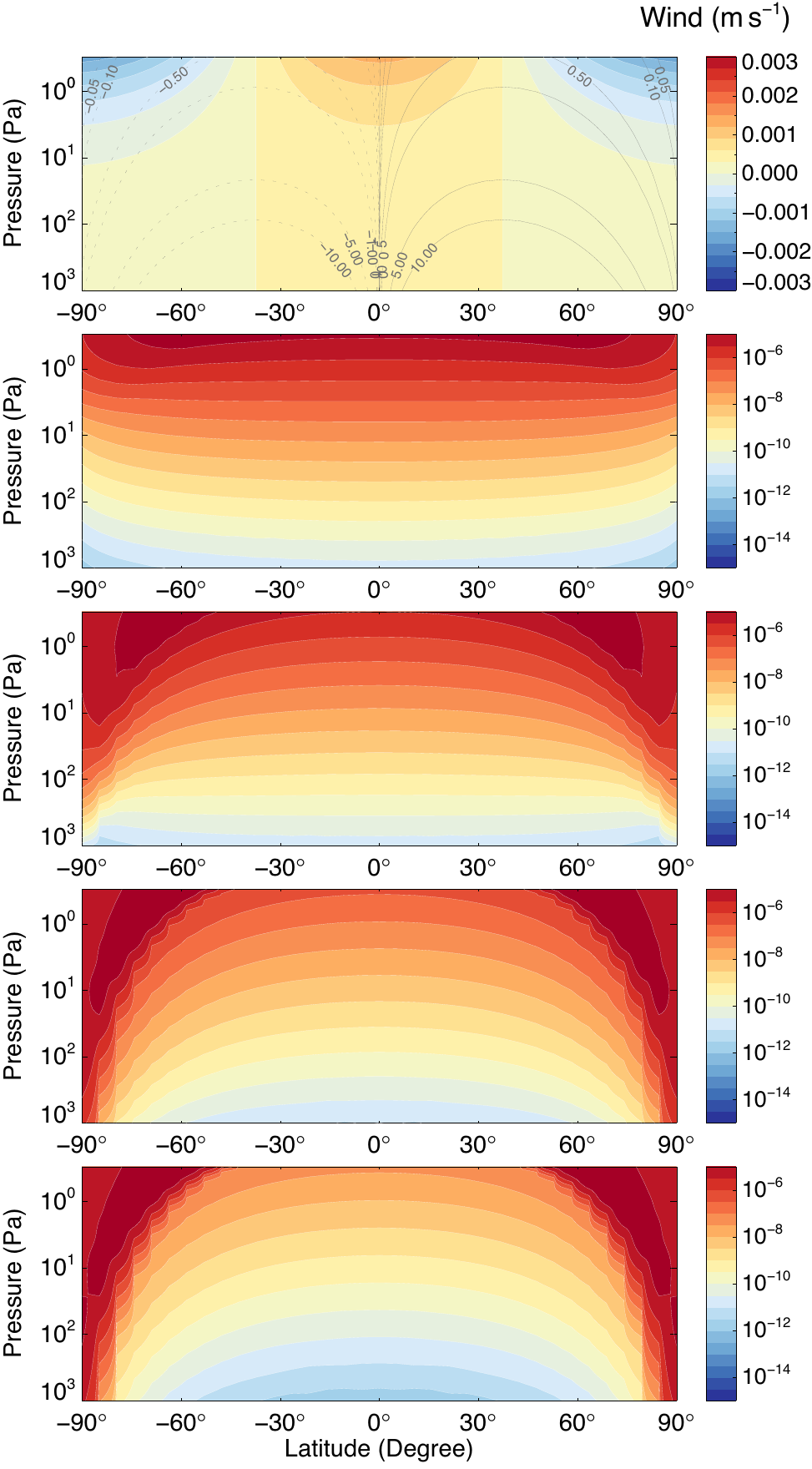} 
 \centering\caption{Latitude-pressure maps of experiments V with non-uniform $\chi_0$. First row: vertical winds (color) and mass streamfunctions (contours, in units of $10^{13}~\mathrm{Kg~s^{-1}}$). Starting from the second row we show volume mixing ratio maps of tracers with chemical timescales of $10^7$ s, $10^8$ s, $10^9$ s and $10^{10}$ s from top to the bottom, respectively.} 
\end{figure}

\begin{figure}[t]
\includegraphics[width=0.48\textwidth]{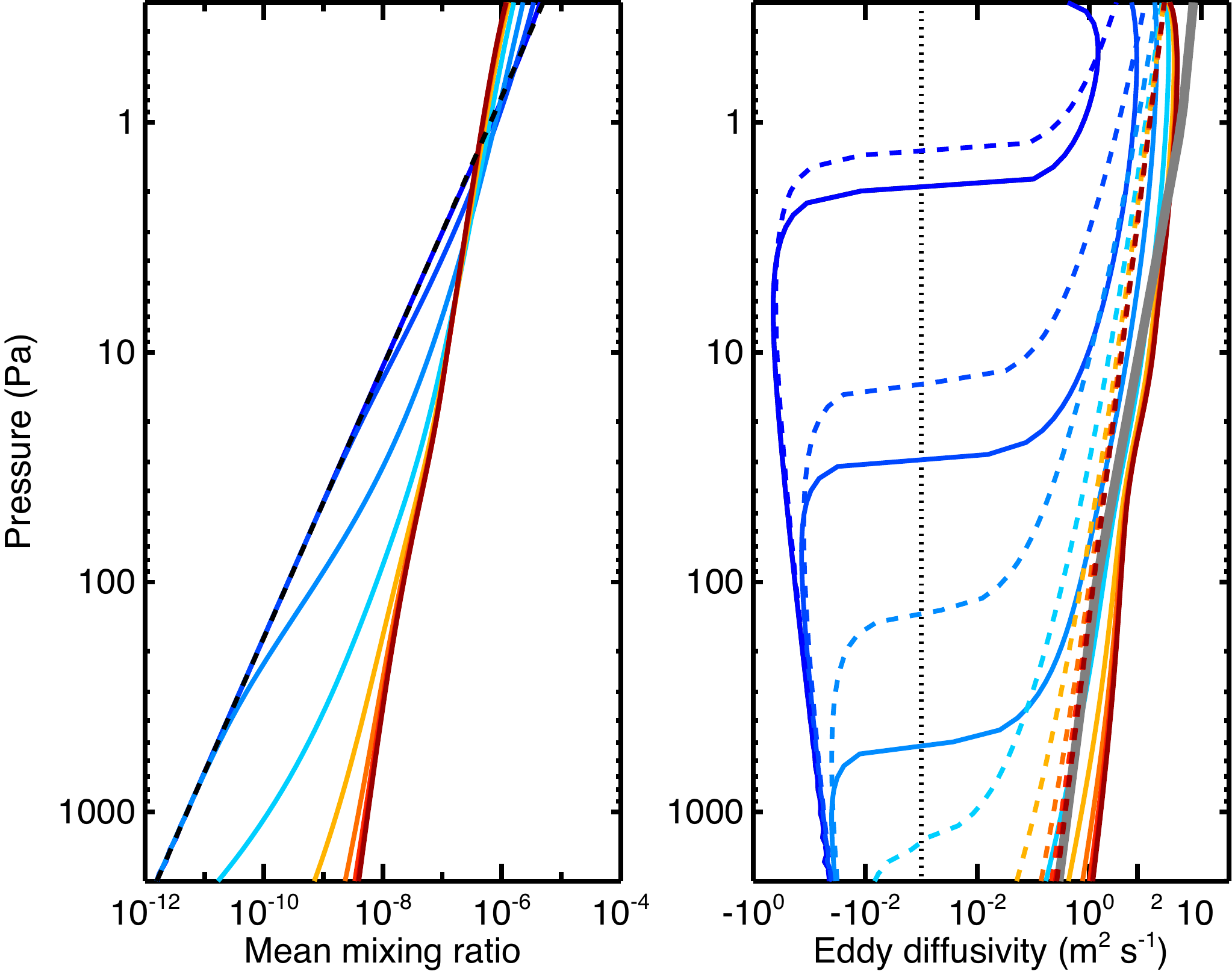} 
  \caption{Same as Fig. 5 but from Experiment V with non-uniform $\chi_0$. The predicted eddy diffusivity profiles based on Eq. (22) with non-diffusive correction are shown in dashed in the right panel.} 
\end{figure}

The cases with non-uniform $\chi_0$ (Regime II) behave quite different from that with uniform $\chi_0$ (Regime I). The numerical simulations in Experiment V are shown in Fig. 10. The prescribed $\chi_0$ follows a cosine function of latitude. The equilibrium mixing ratio is higher at the equator and lower at poles. For the short-lived species, the general patterns of the final tracer distributions roughly follow the distribution of $\chi_0$. But in the atmosphere above 1 Pa where the transport is very efficient, the low-mixing-ratio tracers are advected upward from the lower atmosphere at equator, leading to a smaller tracer abundance at low latitudes than at mid-latitudes. The global-mean tracer mixing ratio profiles roughly follow the global-mean chemical equilibrium tracer profile (Fig. 11). 

As the tracer lifetime increases, the global-mean tracer mixing ratio profile becomes more vertical with pressure due to stronger dynamical mixing (Fig. 11). In those cases, meridional circulation greatly shapes the tracer distributions towards the pattern with lower-mixing-ratio tracers at equator and higher-mixing-ratio at poles (Fig. 10). The latitudinal trend of the tracer on an isobar is opposite to the prescribed chemical equilibrium tracer distribution $\chi_0$. In general the patterns of longer-lived tracers look similar to that in Experiment III, where the tracer source is also from the top atmosphere but $\chi_0$ is flat with latitude in that case. However, there is a significant difference between the Experiments III and V. There are two mid-latitude ``tongues" sinking from the top atmosphere in Experiment V while they are missing in Experiment III, where the tracer mixing ratio is high from the mid-latitudes all the way to the poles. The existence of the mid-latitude tracer maxima in Experiment V suggests strong downwelling of low-mixing-ratio tracers from the top atmosphere in the polar region where the chemical equilibrium tracer mixing ratio is low. The downward tracer fluxes dilute the high-mixing-ratio tracers that are transported from the middle latitudes to the polar region, resulting local maxima (``tongues") that are concentrated at mid-latitudes.

Our theory in Section 2 predicts that non-uniform $\chi_0$ could introduce a negative component in the global-mean eddy diffusivity $K_{zz}$. This effect is more pronounced for short-lived species. The numerically calculated $K_{zz}$ are shown in Fig. 11. The derived $K_{zz}$ become negative below some pressure levels for short-lived species with chemical lifetime smaller than $\mathrm{10^8}$ s. This non-diffusive effect is primarily due to the non-uniform $\chi_0$, which is less important when the tracer lifetime becomes longer and when the tracer distribution substantially deviates away from the chemical equilibrium. For long-live species, the non-diffusive effect vanishes, and the derived $K_{zz}$ profile generally agrees with that from Experiments I, II and III with uniform $\chi_0$. Using our analytical formula of $K_{zz}$ with a non-diffusive correction term (Eq. 22), we can generally reproduced the negative $K_{zz}$ values for short-lived species as well as the positive values for long-lived species (Fig. 11). 

\section{Conclusion and Discussion}

The central assumption in the 1D framework is that vertical tracer transport acts in a diffusive manner in a global-mean sense. Eddy diffusivity is a key parameter in this framework and is normally constrained by fitting the model to observed tracer profiles. However, the physical meaning of this empirically determined quantity is usually not elucidated, and there have been few attempts in the planetary literature to estimate it from first principles or show systematically how it should vary from planet to planet. In this study we investigated some of the fundamental processes that are lumped into---and control---this single quantity. We generalized the pioneering theoretical work from \citet{holton1986dynamically} for a 2D Earth model to a 3D atmosphere and explicitly derived the diffusivity expression from first principles for specific situations. We performed tracer transport simulations in a 2D chemical-diffusion-advection model for rapid-rotating planets using a simple chemical source/sink scheme. By deriving the 1D eddy diffusivity from the globally averaged vertical transport flux, we showed that the simulation results in 2D agree with our theoretical predictions. Therefore this study can serve as a theoretical foundation for future work on estimating or understanding the effective eddy diffusivity for global-mean vertical tracer transport.

The general take-home message from our investigation is that interaction between the chemistry and dynamics is important in controlling the 1D vertical tracer transport. Our work demonstrates that the global-mean vertical tracer transport crucially depends on the correlations between the vertical velocity field and the tracer horizontal variations, which is significantly modulated by atmospheric circulation, horizontal diffusion and wave mixing, and the local tracer chemistry or microphysics. Importantly, this correlation is also controlled by the chemistry itself, which implies that---even for a given atmospheric circulation---the vertical mixing rates and effective eddy diffusivity can differ from one chemical species to another. In general, we found that the traditional assumption in current 1D models that all chemical species are transported via the same eddy diffusivity breaks down. Instead, the 1D eddy diffusivity should increase with tracer chemical lifetime and circulation strength but decrease with horizontal mixing efficiency due to breaking of Rossby waves or other horizontal mixing processes. Our analytical theory including these effects can explain the 2D numerical simulation results over a wide parameter space. This physically motivated formulation of eddy diffusivity could be useful for future 1D tracer transport models.

We emphasize that the conventional ``diffusive" assumption of the global-mean vertical tracer transport does not always hold. In this study we demarcated three regimes in terms of the tracer lifetime and horizontal tracer distribution under chemical equilibrium (Fig. 3). Only in regime I, a short-lived tracer with uniform distribution of chemical equilibrium abundance, is the traditional diffusive assumption mostly valid. In the other two regimes, tracer with non-uniform distribution of chemical equilibrium abundance (regime II) and the tracer lifetime significantly long compared with the transport timescale (regime III), the global-mean vertical tracer transport could be largely influenced by non-diffusive effects. Non-diffusive effects could result in a negative effective eddy diffusivity in the traditional diffusive framework, either for the short-lived species in regime II or for the long-lived species in regime III. A negative diffusivity does not physically make sense and might be difficult to incorporate into 1D models. For relatively long-lived species in regime II (but not long enough to reach regime III), we provided a simple derivation to capture the non-diffusive effects, which might be useful for future 1D models.

Our detailed analytical and numerical analysis concludes several key points:

(1) Larger characteristic vertical velocities contribute to a larger global-mean tracer mixing. Because vertical velocities tend to increase with height, we find that the eddy diffusivity generally increases with height as well, although vertical variations of chemical timescale could complicate this picture if they are sufficiently large. For short-lived tracers in regime I, $K_{zz}\propto\hat{w}^2$ and for the long-lived tracers in all regimes, $K_{zz}\propto\hat{w}$ if the non-diffusive effect is not significant.

(2) Efficient horizontal eddy mixing due to breaking of Rossby waves or other wave processes will smooth out the horizontal variations of tracer and thus decrease the global-mean eddy diffusivity. But the horizontal tracer advection due to the mean flow can either increase or decrease the eddy diffusivity, depending on the eddy tracer flux convergence/divergence induced by the mean flow, which fundamentally depends on the correlation between the horizontal mean flow and the horizontal tracer variations.

(3) Global-mean eddy diffusivity depends on the tracer sources and sinks due to chemistry and microphysics. In an idealized case with linear chemical relaxation, we showed that the effective eddy diffusivity increases with the chemical relaxation timescale (i.e., chemical lifetime). In regime I, $K_{zz}\propto\tau_c$. When the chemical lifetime is very long (regime III), the effective eddy diffusivity reaches its asymptotic value--a product of vertical wind velocity and vertical transport length scale.

(4) In regime I, short-lived species exhibit a similar spatial pattern as the vertical velocity field (as viewed on an isobar). But if the equilibrium chemical field has horizontal variations (regime II), the resulting tracer distribution is complicated. In regime III, the pattern of a long-lived tracer is largely controlled by the atmospheric dynamics. In this case, the correlation between vertical velocity and tracer fields (on isobars) breaks down, and the vertical profile of the tracer abundance differs substantially from its chemical equilibrium profile. The horizontal variation of the long-lived tracers is generally smaller than that of the short-lived tracers due to horizontal mixing. If the tracer lifetime is sufficiently long (i.e., $\tau_c\rightarrow \infty$), the horizontal tracer field is expected to be completely homogenized across the globe, but we did not simulate this physical limit in this study.

(5) The diffusive assumption is generally valid in regime I and the effective eddy diffusivity is always positive using our idealized chemical schemes. But in regime II, there is a strong variation in the equilibrium chemical field. Non-diffusive effects are important in regime II if there is a good correlation between the equilibrium tracer field and the vertical velocity field. In some situations, tracers could be vertically mixed towards the source in a global-mean sense, leading to a negative eddy diffusivity. As the chemical lifetime increases in regime II (but not long enough to reach regime III), the non-diffusive effect could become less significant. 

(6) Non-diffusive behavior could occur in regime III when the tracer chemical lifetime is much longer than the atmospheric dynamical timescale. In this regime, the tracer material surface is distorted significantly and the global-mean tracer profile is complicated. For example, in the case where the tracer source is in the deep atmosphere, the global-mean tracer profile could exhibit a local minimum in the middle atmospheric layers. The derived effective eddy diffusivity will be negative above the local minimum, suggesting a strong non-diffusive effect.

(7) We derived the analytical species-dependent eddy diffusivity for tracers with uniform chemical equilibrium (Eq. 13). We also provided a simple non-diffusive correction for long-lived tracers with non-uniform chemical equilibrium (Eq. 15). The dynamical timescale in our theory depends on the horizontal length scale $L_h$ and/or vertical length scale $L_v$ of the tracer transport. Using the pressure scale height $H$ as $L_v$, the theoretical predictions generally agree with our 2D numerical simulations. For long-lived species, the actual $L_v$ appears larger than $H$ in our 2D simulations.
 
(8) A widely accepted assumption in current 1-D chemical models of planetary atmospheres---Òspecies-independent eddy diffusionÓ, which assumes a single profile of vertical eddy diffusivity for all species---is generally invalid. Using a species-dependent eddy diffusivity in our theory will lead to a more realistic understanding of global-mean tracer transport in planetary atmospheres. 

In this paper, we only focused on fast-rotating planets where a zonal-symmetric assumption generally holds true. But for planetary atmospheres with significant zonal asymmetry, such as the atmospheres on tidally locked exoplanets, a 3D tracer-transport model is necessary to test our theory derived in this study. In a consecutive paper (Paper II, Zhang $\&$ Showman 2018b), we will focus on the tracer transport on tidally locked planets using a GCM and simple tracer schemes and demonstrate that our analytical theory can be also applied to that regime. 

We should note that, in this study we only considered idealized chemical schemes that relax the tracer distribution toward a specified chemical equilibrium over a specified timescale. Other types of sources/sinks, for instance, due to cloud particle setting or due to more complex gas-phase chemistry have yet to be explored. The global-mean vertical transport behaviors of those tracers merit investigation in the future. Future work could also study the seasonal effect of the tracer transport for high-obliquity planets where the circulation patterns significantly change with time.

In our simulations we prescribed the meridional circulation and eddy diffusivities to study the response of the tracer distributions, but this approach prevents us from investigating the detailed dynamical interaction between the waves/eddies and the tracers. Waves and eddies also have strong dynamical effects on zonal jets that influence the tracer transport. For example, the polar jets driven by waves and eddies on Earth could act as potential vorticity barriers for tracer transport (\citealt{cammas1998atlantic}). A 3D GCM relevant for stratospheres of fast-rotating planets is needed in the future to explore the effects of small-to-regional-scale waves and eddies on tracer transport. Furthermore, besides the large-scale upwelling and downwelling transport, waves might also contribute to the vertical mixing in the stratified atmosphere. In the high atmosphere such as the mesosphere on Earth or thermosphere on Jupiter, gravity waves could lead to a strong tracer transport and tracer mixing. \citet{lindzen1981turbulence} first proposed that the gravity wave breaking including the breaking of tidal waves will significantly mix the tracers vertically. \citet{strobel1981parameterization} and \citet{schoeberl1984nonzonal} proposed a parameterization including both gravity wave breaking and tracer mixing by linear waves. \citet{strobel1987vertical} also found that the effective vertical eddy diffusivity due to gravity waves depends on the tracer chemical lifetime. It is expected that the total vertical eddy diffusivity would be a sum of both large-scale mixing and gravity wave mixing. Further investigation is needed so that 1D vertical tracer transport can be understood in a coherent 3D chemical-transport framework from the deep, turbulent atmosphere to the upper, low-density atmosphere including convective mixing, large-scale circulation, and mixing from atmospheric wave processes such as breaking of Rossby waves and gravity waves. 

\appendix
\numberwithin{equation}{section}

\section{Derivation of 1D Effective Eddy Diffusivity in a 2D Chemical-Advective-Diffusive System } 

The governing tracer transport equation of a 2D chemical-advection-diffusive system using the coordinates of log-pressure and latitude is (Eq. 19 in the text, \citealt{shia1989sensitivity}, \citealt{shia1990two}, \citealt{zhang2013jovian}): 
\begin{equation}
\begin{split}
\frac{\partial \chi}{\partial t}+v^*\frac{\partial \chi}{\partial y}+w^*\frac{\partial \chi}{\partial z}-\frac{1}{\cos\phi}\frac{\partial}{\partial y}(\cos\phi K_{yy2D}\frac{\partial\chi}{\partial y})
-e^{z/H}\frac{\partial}{\partial z}(e^{-z/H}K_{zz2D}\frac{\partial\chi}{\partial z})=S
\end{split}
\end{equation}
where $\chi$ is the tracer mixing ratio, $\phi=y/a$ is latitude and $a$ is the planetary radius. $z \equiv -H\log (p/p_0)$ is the log-pressure coordinate, where $H$ is a constant reference scale height, $p$ is pressure and $p_0$ is the pressure at the bottom boundary. $K_{yy2D}$ and $K_{zz2D}$ are the horizontal and vertical eddy diffusivities, respectively. Residual circulation velocities are $v^*$ and $w^*$ in the meridional and vertical directions, respectively. A mass streamfunction $\psi$ can be introduced such that:
 \begin{subequations}
\begin{align}
v^*&=-\frac{1}{2\pi a\rho_0\cos\phi}e^{z/H}\frac{\partial}{\partial z}(e^{-z/H}\psi)
\\
w^*&=\frac{1}{2\pi a\rho_0\cos\phi}\frac{\partial \psi}{\partial y}
\end{align}
\end{subequations}
where $\rho_0$ is the reference density of the atmosphere at log-pressure $z=0$. The $v^*$ and $w^*$ given in Eq. (A.2) naturally satisfy the continuity equation: 
\begin{eqnarray}
\frac{1}{\cos\phi}\frac{\partial}{\partial y}(\cos\phi~v^*)+e^{z/H}\frac{\partial}{\partial z}(e^{-z/H}w^*)=0.
\end{eqnarray}

Combining Eq. (A.1) and Eq. (A.3), we obtain the flux form of the tracer transport equation:
\begin{eqnarray}
\frac{\partial \chi}{\partial t}+\frac{1}{\cos\phi}\frac{\partial}{\partial y}(\cos\phi~v^*\chi)+e^{z/H}\frac{\partial}{\partial z}(e^{-z/H}w^*\chi)-\frac{1}{\cos\phi}\frac{\partial}{\partial y}(\cos\phi K_{yy2D}\frac{\partial\chi}{\partial y})
-e^{z/H}\frac{\partial}{\partial z}(e^{-z/H}K_{zz2D}\frac{\partial\chi}{\partial z})=S.
\end{eqnarray}

Here we also define the eddy-mean decomposition $A=\overline{A}+A^\prime$ where $\overline{A}$ is the latitudinally averaged quantity at constant pressure and $A^\prime$ is the eddy term. Taking the latitudinal average of Eq. (A.3) to eliminate the meridional transport term (the $v^*$ term), we obtain $\overline{w^*}=0$ at isobars if we assume the latitudinal-mean vertical velocity $\overline{w^*}$ vanishes at top and bottom boundaries. We average Eq. (A.4) and both meridional advection and diffusion terms vanish. Using $\overline{w^*}=0$, we obtain:
\begin{eqnarray}
\frac{\partial\overline{\chi}}{\partial t}+e^{z/H}\frac{\partial}{\partial z}(e^{-z/H}\overline{w^*\chi^\prime})-e^{z/H}\frac{\partial}{\partial z}(e^{-z/H}K_{zz2D}\frac{\partial\overline{\chi}}{\partial z})=\overline{S}.
\end{eqnarray}

Based on the ``flux-gradient relationship'', we can introduce an eddy diffusivity $K_w$ to approximate the vertical transport term associated with $w^*$ (the second term in the left hand side) such that:
\begin{eqnarray}
\overline{w^*\chi^\prime}\approx -K_w\frac{\partial\overline{\chi}}{\partial z}.
\end{eqnarray}

Then the 1D global-mean tracer transport equation (Eq. A.5) can be formulated as a vertical diffusion equation: 
\begin{eqnarray}
\frac{\partial\overline{\chi}}{\partial t}-e^{z/H}\frac{\partial}{\partial z}(e^{-z/H}K_{zz}\frac{\partial\overline{\chi}}{\partial z})=\overline{S}
\end{eqnarray} 
where define the total 1D effective eddy diffusivity $K_{zz}=K_w+K_{zz2D}$. Therefore, the $K_{zz2D}$ vertical eddy diffusion term in Eq. (A.5) can be treated as an additive term in the total global-mean effective eddy diffusivity $K_{zz}$. 

In order to analytically derive $K_w$, we subtract both Eq. (A.5) and Eq. (A.3) multiplied by $\overline{\chi}$ from Eq. (A.4):
\begin{equation}
\frac{\partial\chi^\prime}{\partial t}+\frac{1}{\cos\phi}\frac{\partial}{\partial y}(\cos\phi~v^*\chi^\prime)-\frac{1}{\cos\phi}\frac{\partial}{\partial y}(\cos\phi K_{yy2D}\frac{\partial\chi^\prime}{\partial y})+w^*\frac{\partial\overline{\chi}}{\partial z}+e^{z/H}\frac{\partial}{\partial z}[e^{-z/H}(w\chi^\prime-\overline{w^*\chi^\prime}-K_{zz2D}\frac{\partial\chi^\prime}{\partial z})]=S^\prime.
\end{equation}

To solve this equation for $\chi^\prime$, we made four assumptions that are similar to those in Section 2.3.

(i) We neglect the first temporal evolution term in statistical steady state. 

(ii) We neglect the complicated eddy term (the fifth term) by assuming the vertical transport of the tracer eddy $\chi^\prime$ is much smaller than that of the latitudinal-mean tracer ($\overline{\chi}$, the fourth term). Here we have also assumed the vertical diffusion of $\chi^\prime$ (the $K_{zz2D}$ term) is much smaller than than the vertical advection of the mean tracer.  

(iii) We approximate the horizontal tracer eddy flux terms (second and third terms) in a linear relaxation form. In Section 2.3, we have decomposed the horizontal mixing processes in a 3D atmosphere into the tracer advection by zonal-mean flow and the tracer diffusion by horizontal eddies and waves. In the zonal-mean 2D system, the second and third terms in the left hand side of Eq. (A.8) correspond to the advection and diffusion process, respectively. For the advection term, we have:
\begin{equation}
\frac{1}{\cos\phi}\frac{\partial}{\partial y}(\cos\phi~v^*\chi^\prime)\approx \frac{\chi^\prime}{\tau_{adv}}
\end{equation}
where $\tau_{adv}\approx L_h/\hat{v^*}$ where $L_h$ is the horizontal length scale and $\hat{v^*}$ is the meridional velocity scale. From the continuity equation (Eq. A.3), we can also show $\tau_{adv}\approx L_v/\hat{w^*}$, where $L_v$ is the vertical length scale and $\hat{w^*}$ is the vertical velocity scale. Similarly, the diffusive term can be approximated as:
\begin{equation}
-\frac{1}{\cos\phi}\frac{\partial}{\partial y}(\cos\phi K_{yy2D}\frac{\partial\chi^\prime}{\partial y})\approx \frac{\chi^\prime}{\tau_{diff}}
\end{equation}
where $\tau_{diff}\approx a^2/K_{yy2D}$. Here we have made an additional assumption that the positive tracer anomaly ($\chi^\prime>0$) is associated with the tracer eddy flux divergence (i.e., the sign of the left hand side is positive in this situation) and the negative anomaly is associated with the tracer eddy flux convergence. This is generally a good assumption as long as the material surface of the tracer is not very complicated. For example, if the horizontal tracer distribution can be approximated by a second-degree Legendre polynomial as in \citet{holton1986dynamically}, one can show that our above argument holds true.

(iv) We also consider a linear chemical scheme which relaxes the tracer distribution towards local chemical equilibrium $\chi_0$ in a timescale $\tau_c$. After the mean-eddy decomposition, we obtain the latitudinal mean of the chemical source/sink term $\overline{S} = (\overline{\chi_0}-\overline\chi)/\tau_c$ and the departure $S^\prime=(\chi_0^{\prime}-\chi^\prime)/\tau_c$. See Section 2.3 for details.

With above assumptions, Eq. (A.8) can be simplified as:
\begin{equation}
w^*\frac{\partial\overline{\chi}}{\partial z}+\frac{\chi^\prime}{\tau_{adv}}+\frac{\chi^\prime}{\tau_{diff}}=\frac{\chi_0^{\prime}-\chi^\prime}{\tau_{c}}.
\end{equation}

We solve for $\chi^\prime$:
\begin{equation}
\chi^\prime=\frac{-w^*\frac{\partial\overline{\chi}}{\partial z}+\tau_{c}^{-1}\chi_0^{\prime}}{\tau_{diff}^{-1}+\tau_{adv}^{-1}+\tau_{c}^{-1}}.
\end{equation}
Inserting Eq. (A.12) in Eq. (A.6), we solve for $K_w$ and finally obtain the analytical expression for $K_{zz}$: 
\begin{equation}
K_{zz}\approx K_{zz2D}+\frac{\overline{w^{*2}}}{\tau_{diff}^{-1}+\tau_{adv}^{-1}+\tau_{c}^{-1}}-\frac{\overline{w^*\chi_0^{\prime}}}{1+\tau_{diff}^{-1}\tau_{c}+\tau_{adv}^{-1}\tau_{c}}(\frac{\partial\overline{\chi}}{\partial z})^{-1}.
\end{equation}

In our 2D cases, if the chemical equilibrium abundance $\chi_0$ is not uniformly distributed across the latitude, the magnitude of $\chi_0^{\prime}$, $\Delta{\chi_0^{\prime}}$, can be approximated by the root-mean-square of $\chi_0^{\prime}$ over the globe. One can show that, for a cosine function of $\chi_0$, $\Delta{\chi_0^{\prime}} \approx 0.28~\overline{\chi_0}$ from the area-weighted global integration. The vertical gradient of $\overline{\chi}$ can be approximated by that of the global-mean chemical equilibrium mixing ratio $\overline{\chi_0}$. Given the specific meridional circulation patterns in Eq. (20), the area-weighted global-mean vertical velocity scale is $\hat{w^*}=\gamma w_0e^{\eta z/H}$. If the tracers are transported by global-scale circulation patterns, we can assume that the horizontal transport length scale $L_h$ is the planetary radius $a$ and vertical transport length scale $L_v$ is the pressure scale height $H$, we predict $K_{zz}$ for our 2D cases (Eq. 22 in the text): 
\begin{equation}
K_{zz}\approx K_{zz2D}+\frac{\gamma^2w_0^2e^{2\eta z/H}}{K_{yy2D} a^{-2}+\gamma w_0e^{\eta z/H} H^{-1}+\tau_{c}^{-1}}-\frac{\gamma w_0e^{\eta z/H}\Delta{\chi_0^{\prime}}}{1+\tau_{c}K_{yy2D} a^{-2}+\tau_{c}\gamma w_0e^{\eta z/H} H^{-1}}(\frac{\partial\overline{\chi_0}}{\partial z})^{-1}.
\end{equation}

If the chemical equilibrium abundance $\chi_0$ is uniformly distributed across the latitude, $\chi_0^{\prime}$ is zero and thus the last term in the right hand side of Eq. (A.13) vanishes. $K_{zz}$ becomes (Eq. 23 in the text): 
\begin{equation}
K_{zz}\approx K_{zz2D}+\frac{\gamma^2w_0^2e^{2\eta z/H}}{K_{yy2D} a^{-2}+\gamma w_0e^{\eta z/H} H^{-1}+\tau_{c}^{-1}}.
\end{equation}

\section{Acknowledgements}		
This research was supported by NASA Solar System Workings Grant NNX16AG08G to X.Z. and A.P.S.. We dedicate this work to Dr. Mark Allen (1949-2016), one of the founders of the Caltech/JPL kinetics model. We thank Y. Yung, R. Shia, and J. Moses for providing the eddy diffusion profiles in their chemical models. Some of the simulations were performed on the Stampede supercomputer at TACC through an allocation by XSEDE.

\bibliographystyle{apj}

\end{document}